\documentclass[12pt, draftclsnofoot, onecolumn]{IEEEtran}
\usepackage{amssymb}
\usepackage{amsmath}
\usepackage{cite}
\usepackage{url}
\usepackage{xcolor}
\usepackage{cite,graphicx,amsmath,amssymb}
\usepackage{subfigure}
\usepackage{citesort}
\usepackage{fancyhdr}
\usepackage{mdwmath}
\usepackage{mdwtab}
\usepackage{caption}
\usepackage{amsthm}
\usepackage{setspace}
\usepackage{multirow}
\usepackage{epstopdf}
\usepackage[justification=centering]{caption}
\newtheorem{remark}{Remark}
\newtheorem{theorem}{Theorem}

\newtheorem{lemma}{Lemma}

\newtheorem{corollary}{Corollary}

\usepackage{threeparttable}
\usepackage{algorithm}
\usepackage{algorithmic}
\usepackage{cases}

\begin{document}
	
	\title{Exploiting NOMA and RIS in Integrated Sensing and Communication 	\vspace{-0.2cm}} 	

	\author{
		Jiakuo~Zuo,
		Yuanwei~Liu,~\IEEEmembership{Senior Member,~IEEE},
		         Chenming Zhu, Yixuan Zou,~\IEEEmembership{Graduate Student Member,~IEEE}, Dengyin Zhang and Naofal Al-Dhahir,~\IEEEmembership{Fellow,~IEEE}\vspace{-0.9cm}
		 \thanks{Part of this work has been accepted by the IEEE GLOBECOM Workshop on NGMA for Future Wireless Communications, Rio de Janeiro, Brazil, 4–8 Dec. 2022~\cite{RISNOMAISAC}.} 
		\thanks{J. Zuo is with the Institute of Communications and Information Technology, China Information Consulting and Designing Institute Company Ltd., Nanjing 210003, China, and also with School of Internet of Things, Nanjing University of Posts and Telecommunications, Nanjing, China. (email:zuojiakuo@njupt.edu.cn).}
		 \thanks{Y. Liu and Y. Zou are with the School of Electronic Engineering and Computer Science, Queen Mary University of London, London E1 4NS, U.K. (e-mail:yuanwei.liu@qmul.ac.uk, yixuan.zou@qmul.ac.uk ).}
	   \thanks{C. Zhu is with the Institute of Communications and Information Technology, China Information Consulting and Designing Institute Company Ltd., Nanjing 210003, China. (email:e-mail: zhuchenming@cicdi.com).}
	   \thanks{D. Zhang is with School of Internet of Things, Nanjing University of Posts and Telecommunications, Nanjing, China. (email:zhangdy@njupt.edu.cn).}
	   \thanks{N. Al-Dhahir is with the Department of Electrical and Computer Engineering, The University of Texas at Dallas, Richardson, TX 75080 USA. (email:aldhahir@utdallas.edu).}
	   }
	\maketitle
	\vspace{-1cm}
  \begin{abstract}
  	 	\vspace{-0.3cm}
 	  \textcolor[rgb]{0.00,0.00,0.00}{A novel integrated sensing and communication (ISAC) system is proposed, where a dual-functional base station is utilized to transmit the superimposed non-orthogonal multiple access (NOMA) communication signal for serving communication users and sensing targets simultaneously. Furthermore, a new reconfigurable intelligent surface (RIS)-\textcolor[rgb]{0.00,0.00,0.00}{aided}-sensing structure, where \textcolor[rgb]{0.00,0.00,0.00}{a} dedicated RIS is deployed to provide virtual line-of-sight (LoS) links for radar targets, is also proposed to address the significant path loss or blockage of LoS links for \textcolor[rgb]{0.00,0.00,0.00}{the} sensing task. Based on this setup, the beampattern gain at the RIS for \textcolor[rgb]{0.00,0.00,0.00}{the} radar target is derived and adopted \textcolor[rgb]{0.00,0.00,0.00}{as a} sensing metric. }
    The objective of this paper is to maximize the minimum beampattern gain by jointly optimizing active beamforming at \textcolor[rgb]{0.00,0.00,0.00}{the} base station (BS), power allocation coefficients among NOMA users and passive beamforming at the RIS. 
  \textcolor[rgb]{0.00,0.00,0.00}{To tackle the non-convexity of the formulated optimization problem, the beampattern gain and constraints are first transformed into more tractable forms. Then, an iterative block coordinate descent (IBCD) algorithm is proposed by employing successive convex approximation (SCA), Schur complement, semidefinite relaxation (SDR) and sequential rank-one constraint relaxation (SRCR) methods.}
 To reduce the complexity of the proposed IBCD algorithm, a low-complexity iterative alternating optimization (IAO) algorithm is proposed. Particularly, the active beamforming is optimized by solving a semidefinite programming (SDP) problem and the closed-form solutions of the power allocation coefficients are derived. Numerical results show that: \textcolor[rgb]{0.00,0.00,0.00}{i) the proposed RIS-NOMA-ISAC system always outperforms the RIS-ISAC system without NOMA in beampattern gain and illumination power; ii) the low-complexity IAO algorithm achieves a comparable performance to that achieved by the IBCD algorithm. iii) high beampattern gain can be achieved by the proposed joint optimization algorithms in underloaded and overloaded communication scenarios. }
 	\vspace{-0.5cm}
 \end{abstract}
 
 \begin{IEEEkeywords}
 	\vspace{-0.5cm}
 	Beamforming optimization, integrated sensing and communication, non-orthogonal multiple access, reconfigurable intelligent surface.
 \end{IEEEkeywords}
  	\vspace{-0.5cm}
\section{Introduction}
For conventional wireless communication systems, the main \textcolor[rgb]{0.00,0.00,0.00}{performance} indicator is the communication quality of service (QoS). However, in the upcoming beyond fifth generation (B5G) and sixth generation (6G) wireless networks, sensing service will play a more important role than ever before~\cite{9585321,Dong2022Sensing}. The goal of future wireless communication is to provide services based on both sensing and communication functionalities. Toward this \textcolor[rgb]{0.00,0.00,0.00}{goal}, integrated sensing and communication (ISAC) has emerged and attracted growing attention in both academia and industries~\cite{9755276,Li2022AssistingLB,9585321}. Broadly, the research contributions on ISAC can be classified into two categories, namely, radar-communication coexistence (RCC) and dual-function radar communication (DFRC)~\cite{Liu2022ISAC}. For RCC, the radar and communication devices are separated, while the spectrum is shared simply by radar and communication systems. Sophisticated interference management and cooperation techniques are needed for RCC to mitigate the system interference. However, this requires \textcolor[rgb]{0.00,0.00,0.00}{a} large number of information \textcolor[rgb]{0.00,0.00,0.00}{exchanges} and leads to high overhead.
For DFRC, the radar and communication functions share the same hardware and a unified transmit waveform is exploited to simultaneously perform communication and sensing tasks. Nevertheless, it is a challenging task to design \textcolor[rgb]{0.00,0.00,0.00}{an} efficient waveform to balance the conflicting requirements between the communication and sensing \textcolor[rgb]{0.00,0.00,0.00}{tasks}.  

Various key techniques, such as millimeter wave (mmWave)~\cite{Gao2022IntegratedSA,9729809,9849103}, terahertz (THz)~\cite{Chaccour2021JointSA,9729746,Ahmet2022}, massive multiple-input
multiple-output (MIMO)~\cite{You2022BeamSI} and multi-unmanned aerial vehicle  (UAV)~\cite{9729746,9858656}, have been incorporated in ISAC \textcolor[rgb]{0.00,0.00,0.00}{systems} to boost the communication and sensing capabilities,  
Given the already large
number of works on ISAC communications, in this
paper we focus on the applications of non-orthogonal multiple access (NOMA) and reconfigurable intelligent surface (RIS) to ISAC (or DFRC) systems.
	\vspace{-0.3cm}
\subsection{Prior Works}
A handful of works have been devoted to studying NOMA-ISAC and RIS-ISAC systems. We elaborate on them as follows.
\subsubsection{Studies on NOMA-ISAC systems}
NOMA has received significant attention as a promising technique for supporting massive connectivity, interference management, and spectrum enhancement~\cite{9693417,9167258}.
The authors of~\cite{Mu2022NOMAAidedJR} first proposed a novel NOMA-aided joint radar and multicast-unicast communication system, where the base station (BS) transmits superimposed multicast and unicast messages to the radar users and communication users, while detecting the radar user targets.
The authors of~\cite{Wang2022NOMAEI} utilized the superimposed NOMA communication signal to perform radar sensing and proposed a beamforming design algorithm to maximize the weighted sum of the communication throughput and the effective sensing power.
Since the sensing signal can also be information-bearing, the authors of~\cite{Wang2021NOMAII} utilized multiple beams of the sensing signal to deliver extra multicast information while detecting radar targets. The communication users received one desired unicast stream and multiple multicast streams, which were detected \textcolor[rgb]{0.00,0.00,0.00}{by successive} interference cancellation (SIC).
In conventional ISAC networks, the total bandwidth was assumed to be utilized for both radar detection and wireless communication. However, this assumption is \textcolor[rgb]{0.00,0.00,0.00}{impractical because the bandwidth is} occupied by different applications. To overcome this difficulty, the authors of~\cite{Zhang2022SemiIntegratedSensingandCommunicatios} proposed an uplink NOMA-assisted semi-ISAC system, where the total bandwidth was divided into three portions, namely the ISAC bandwidth, the communication-only bandwidth, and the sensing-only bandwidth.
\subsubsection{Studies on RIS-ISAC systems}
Although many key enablers, such as mmWave, MIMO and THz, have demonstrated tremendous potential in improving the performances of the ISAC system, they usually suffer from various practical limitations, such as \textcolor[rgb]{0.00,0.00,0.00}{uncontrollable electromagnetic} waves propagation, high hardware complexity and energy consumption~\cite{Liu2022ISAC}. Recently, a novel energy-efficient technique, namely, RIS~\cite{9424177}, \textcolor[rgb]{0.00,0.00,0.00}{has} been proposed as an important enabling technology for enhancing the communication performance. By smartly adjusting the amplitude and phase response of the reflecting elements, RIS can reconfigure the electromagnetic environments. Furthermore, by deploying RIS, additional optimization degrees of freedoms (DoFs) and virtual line-of-sight (LoS) links can be introduced for radar targets. 
In~\cite{Sankar2022BeamformingII}, the authors considered two scenarios \textcolor[rgb]{0.00,0.00,0.00}{for} RIS-ISAC systems. In the first scenario, the users and targets were well separated and a single RIS was deployed to enhance the communication performance. The transmit beamformers and RIS phase shifts were designed by minimizing the total beampattern mismatch error and the average squared cross-correlation. In the second scenario, two geographically separated RISs were deployed for sensing and communications.  The transmit beamformers and RIS phase shifts were optimized by maximizing the worst-case target illumination power.
The authors of~\cite{Hua2022JointAA} studied \textcolor[rgb]{0.00,0.00,0.00}{the} transmit power minimization problem and considered two cases. For case 1, the interference introduced by the RIS was perfectly canceled and the cross correlation design was ignored. 
For case 2, both the interference introduced by the RIS and the cross-correlation pattern design were considered. In addition, \textcolor[rgb]{0.00,0.00,0.00}{the authors proved} that the dedicated radar signals were not required in case 1, while the dedicated radar signals were required in case 2 to enhance the system performance.
The authors of~\cite{Xing2022PassiveBD} leveraged RIS to assist \textcolor[rgb]{0.00,0.00,0.00}{a mmWave} ISAC system and proposed a novel passive beamforming strategy by considering the target size. The closed-form detection probability for target sensing was derived and a new concept of ultimate detection resolution was introduced to measure the target detection capability.
Furthermore, the resource allocation for mmWave ISAC system was studied in~\cite{9838546}, where the authors formulated a sum-rate maximization problem and derived closed-form expressions for the radar signal covariance matrix, the communication beamforming vectors and the RIS phase shifts. 
For THz ISAC system, the authors of~\cite{Liu2022ProximalPO} explored the implementation of RIS to \textcolor[rgb]{0.00,0.00,0.00}{compensate for} the path loss caused by molecular absorption. In addition, a joint beamforming optimization algorithm was proposed,  \textcolor[rgb]{0.00,0.00,0.00}{which utilized} gradient-based, primal-dual proximal policy optimization in the multi-user multiple-input single-output
(MISO) scenario.
In practical scenarios, the LoS link between the BS (or access point) and the sensing targets \textcolor[rgb]{0.00,0.00,0.00}{is} likely to be blocked. As a result, target sensing is not applicable. To resolve this issue, the authors of~\cite{Song2022JointTA,Song2022IntelligentRS} utilized RIS to establish virtual LoS links for target sensing and proposed efficient passive beamforming optimization algorithms to guarantee the targets sensing performance.
However, the prior works were focused on the traditional reflection-only RIS, which requires the users and targets to be located at the same side of the BS and RIS. To overcome this limitation, the authors of~\cite{Wang2022STARSEI} proposed a simultaneously transmitting and reflecting intelligent surface (STARS) enabled ISAC system, where STARS can provide $360^{\mathrm{o}}$ full-space coverage and \textcolor[rgb]{0.00,0.00,0.00}{divide} the whole space into two half-spaces, namely the communication space and the sensing space.
	\vspace{-0.5cm}
\subsection{Motivations and Contributions}
\textcolor[rgb]{0.00,0.00,0.00}{Motivated} by the aforementioned NOMA-ISAC and RIS-ISAC systems, this paper aims to achieve further performance enhancement by investigating the integration of NOMA and RIS in ISAC systems. The RIS-NOMA-ISAC system is still a nascent field and many open issues deserve exploration. Note that the integration of NOMA and RIS in \textcolor[rgb]{0.00,0.00,0.00}{an} ISAC system will provide more DoFs for system \textcolor[rgb]{0.00,0.00,0.00}{design} and optimization, which makes the optimization problem different from that of the NOMA-ISAC system or RIS-ISAC system. The motivations and challenges of this paper \textcolor[rgb]{0.00,0.00,0.00}{can be summarized as follows}.
\begin{itemize}
  \item In future ISAC networks, \textcolor[rgb]{0.00,0.00,0.00}{a} massive number of devices will be connected to the networks. The communication users and radar targets will suffer from severe interference. Although multi-antenna techniques can be employed to mitigate the inter-user interference, they cannot efficiently support both communication and sensing functions due to the limited spatial DoFs, \textcolor[rgb]{0.00,0.00,0.00}{which motivates} us to consider \textcolor[rgb]{0.00,0.00,0.00}{the} NOMA technique.
  By applying SIC, NOMA can reduce the impact of \textcolor[rgb]{0.00,0.00,0.00}{limited spatial DoFs}, by partially removing inter-user interference and supporting more communication users with the same orthogonal resource.
	\item In conventional ISAC systems, radar sensing relies on LoS links between BS and radar targets, while the direct communication links between the BS and users are generally assumed to be available. However, in practical scenarios, the radar targets are likely to be distributed in the non-LoS (NLoS) region of the BS and the direct communication links are often blocked, which \textcolor[rgb]{0.00,0.00,0.00}{motivates us to apply the} RIS technique to provide virtual LoS links and reflection links for radar sensing and communication, respectively.	
	\item By exploiting NOMA and RIS techniques, more DoFs can be introduced to enhance the design flexibility of ISAC systems. However, more optimization variables are introduced and the optimization complexity increases. Specifically, it is still a challenging issue to jointly optimize the active beamforming at the BS, the power allocation among NOMA users, and the passive beamforming at the RIS, while guaranteeing the radar-specific requirements and QoS requirements of communication users.
\end{itemize}
 
Inspired by the aforementioned research, motivations and challenges, this paper focus on the design of
the RIS-NOMA-ISAC system.
To the best of our knowledge, the joint optimization design
of the considered system has not been studied yet. Our contributions are summarized as follows.
\begin{enumerate}
	\item \textcolor[rgb]{0.00,0.00,0.00}{We propose a novel RIS-NOMA-ISAC system with a RIS-\textcolor[rgb]{0.00,0.00,0.00}{aided}-sensing structure, where a dedicated RIS is leveraged to introduce virtual LoS links for target sensing and to establish reflection links for user communication. Based on this, we derive the beampattern gain at the RIS as the sensing metric.} In addition, to achieve high spectral efficiency, we exploit the SIC technique of NOMA to multiplex communication users and to mitigate the inter-user interference. \textcolor[rgb]{0.00,0.00,0.00}{To improve the sensing performance while \textcolor[rgb]{0.00,0.00,0.00}{guaranteeing} the minimum QoS required for the communication users, we formulate the beampattern gain maxmin optimization problem by jointly optimizing active beamforming, power allocation coefficients, and passive beamforming.}
	\item The main challenges of solving the formulated problem is the \textcolor[rgb]{0.00,0.00,0.00}{high} coupling between the optimization variables and the exponential expressions of RIS phase shifts. We propose an iterative block coordinate descent (IBCD) algorithm to solve \textcolor[rgb]{0.00,0.00,0.00}{the formulated optimization problem}. \textcolor[rgb]{0.00,0.00,0.00}{In particular, we first transform the optimization problem into a more tractable form, then we divide the coupled variables into two blocks. Finally, we exploit semidefinite relaxation (SDR), successive convex approximation (SCA)\textcolor[rgb]{0.00,0.00,0.00}{, and the} sequential rank-one constraint relaxation (SRCR) algorithm to solve the resultant problems.	}
	\item To reduce the complexity of the proposed IBCD algorithm, we further develop a low-complexity iterative alternating optimization (IAO) algorithm, where we \textcolor[rgb]{0.00,0.00,0.00}{derive closed-form solutions for} the power allocation coefficients in each NOMA cluster and obtain the active beamforming solution by solving a semidefinite programming (SDP) problem. 
	\item  \textcolor[rgb]{0.00,0.00,0.00}{Our simulation results verify the effectiveness of the proposed algorithms and demonstrate the \textcolor[rgb]{0.00,0.00,0.00}{superior ISAC performance} of the proposed RIS-NOMA-ISAC system over the RIS-ISAC system without NOMA. We also obtain some insights from the simulation results. Firstly, the IBCD algorithm outperforms the IAO algorithm but has a higher complexity. Secondly, the proposed RIS-NOMA-ISAC system can \textcolor[rgb]{0.00,0.00,0.00}{achieve} high sensing performance \textcolor[rgb]{0.00,0.00,0.00}{for} underloaded and overloaded communication scenarios.}
\end{enumerate}
 	\vspace{-0.5cm}
\subsection{Organization}
 The rest of this paper is organized as follows. In Section II, the system model and the problem formulation for designing the RIS-NOMA-ISAC system are presented. In Sections III and IV, we present the proposed IBCD algorithm and IAO algorithm to solve the original optimization problem, respectively. Numerical results are presented in Section V, which is followed by the conclusions in Section VI.

Notations: $\mathbb{C}^{M \times 1}$ denotes the space of 
$M\times 1$ complex valued vectors, diag(\textbf{x}) denotes a diagonal matrix whose diagonal elements correspond to vector \textbf{x}. The $(m,n)$-th element of matrix $\textbf{X}$ is denoted as $\left[ \mathbf{X} \right] _{m,n}$. ${\textbf{x}}^{H}$ and ${\textbf{X}}^{H}$ denote the conjugate transpose of vector \textbf{x} and matrix \textbf{X}, respectively. The notations Tr(\textbf{X}) and rank(\textbf{X}) denote the trace and rank of matrix \textbf{X}, respectively. $\mathcal{C}\mathcal{N}\left( 0,\sigma ^2 \right) $ represents the distribution
of a circularly symmetric complex Gaussian variable (CSCG) with zero mean and $\sigma ^2$ variance.
\vspace{-0.3cm}
 \section{System Model and Problem Formulation}
  \begin{figure}[t]
 	 \setlength{\belowcaptionskip}{-25pt}
 	\centering
 	\includegraphics[scale=0.25]{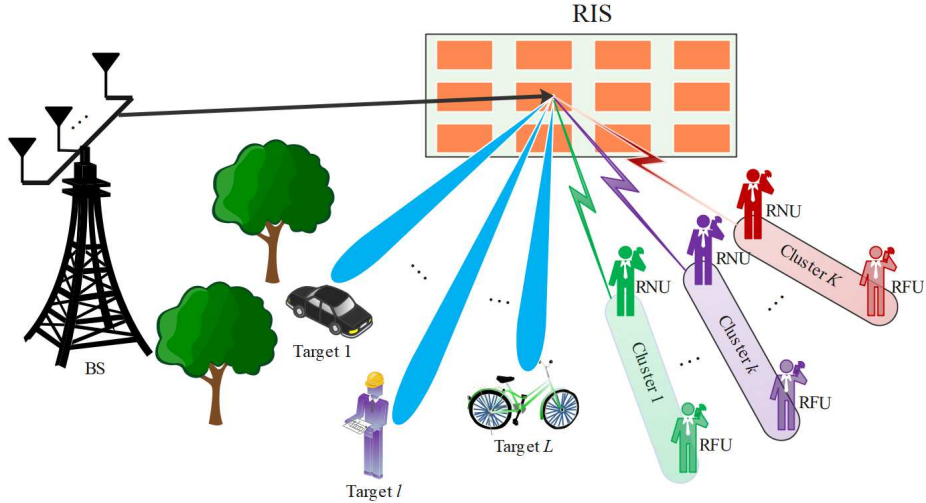}
 	\caption{Illustration of the proposed RIS-NOMA-ISAC system.}
 	\label{ST1}
 \end{figure}
As shown in Fig.~\ref{ST1}, a RIS-NOMA-ISAC system is proposed, which consists of a dual-functional BS equipped with $N$  antennas, $2K$ single-antenna users, \textcolor[rgb]{0.00,0.00,0.00}{a} uniform linear array (ULA)-RIS with $M$ reflecting elements, and $L$ radar targets. We assume that all the direct links from the BS to the communication users and to the radar targets are blocked, which prevents the BS from performing communication and sensing tasks. 
By carefully deploying the RIS, virtual LoS links can be established from the BS to the communication users and to the radar targets. Thus, the performance of communication and sensing tasks can be significantly enhanced. To improve the spectral efficiency and to reduce the system load, we assume that the $2K$ users are grouped into $\emph{K}$ clusters by employing user clustering techniques$\footnote{Many efficient user clustering methods have been proposed for NOMA systems, such as: many-to-one matching~\cite{9167258}, K-means~\cite{8454272}, and correlation of channels~\cite{8485639}. These methods can also be applied for our considered system. In this paper, we only focus on the system optimization problem after user clustering.} $.
As a result, there are two users 
in each cluster, namely the RIS-near user (RNU) and the RIS-far user (RFU). Generally, the
RNUs are much closer to the RIS than the RFUs.  In each
cluster, \textcolor[rgb]{0.00,0.00,0.00}{the} NOMA protocol is applied to serve all users with the same orthogonal resource.
The cluster and user sets are denoted by $\mathcal{K} =\left\{ 1,\cdots ,K \right\}$ and $\mathcal{U} =\left\{ 1,\cdots ,2K \right\}$, respectively. Moreover, ${{\mathcal U}_k}$ denotes the set of users in cluster $k$, where $\mathcal{U} =\cup _{k\in\mathcal{K}} \mathcal{U}  _k$, $ \mathcal{U}  _k\cap  \mathcal{U}  _{\underline{k}}=\oslash \left( k,\underline{k}\in  \mathcal{K} ,k\ne \underline{k} \right) $. For notation simplicity, we denote the RNU and RFU in the $k$-th cluster as \textcolor[rgb]{0.00,0.00,0.00}{users} $\mathrm{U}\left( k,n \right) $ and $\mathrm{U}\left( k,f \right) $, respectively. 
	\vspace{-0.5cm}
\subsection{Communication Model}
The superimposed communication signal transmitted by the BS is given by  
\begin{equation}\label{superimposed_signal}
  \setlength{\abovedisplayskip}{5pt}
 \setlength{\belowdisplayskip}{5pt}
	\mathbf{x}=\sum_{k=1}^K{\mathbf{w}_k\left( \sqrt{a _{k,n}}s_{k,n}+\sqrt{a _{k,f}}s_{k,f} \right)}, 
\end{equation}
where $\mathbf{w}_k\in \mathbb{C}^{N\times 1}$ is the active beamforming vector of the $k$-th cluster, $s_{k,i}$ denotes the communication signal to be sent to user $\mathrm{U}\left( k,i \right) $ with $\mathbb{E}\left( s_{k,i}^{H}s_{k,i} \right) =1$, and $a _{k,i}$ is the corresponding power allocation coefficient, $i\in \left\{ n,f \right\} $, $k\in \mathcal{K}$.

Let $\mathbf{G}\in \mathbb{C}^{M\times N}$ and $\mathbf{g}_{k,i}\in \mathbb{C}^{M\times 1}$ be the channel coefficients of the communication links $\text{BS}\rightarrow \text{RIS}$ and $\text{RIS}\rightarrow \mathrm{U}\left( k,i \right)$, respectively. Furthermore, let $\mathbf{v}=\left[ e^{j\theta _{1}^{\text{RIS}}}e^{j\theta _{2}^{\text{RIS}}}\cdots e^{j\theta _{M}^{\text{RIS}}} \right] 
$ be the passive beamforming vector with $\theta _{m}^{\text{RIS}}\in \left[ 0,2\pi \right) $ denoting the phase shift of the $m$-th reflecting element, $m\in\mathcal{M}\triangleq \left\{ 1,2,\cdots , M \right\} $. With the help of RIS, the signal received at user $\mathrm{U}\left( k,i \right) $ can be mathematically expressed as
 \begin{equation}\label{signal_yki}
    \setlength{\abovedisplayskip}{5pt}
 	\setlength{\belowdisplayskip}{5pt}
 	\begin{split}
		y_{k,i} = 
		& \underset{\text{desired}~ \text{signal}}{\underbrace{\left( \mathbf{g}_{k,i}^{H}\mathbf{\Theta G} \right) \mathbf{w}_k\sqrt{a _{k,i}}s_{k,i}}}+\underset{\text{intra}-\text{cluster}~ \text{interference}}{\underbrace{\left( \mathbf{g}_{k,i}^{H}\mathbf{\Theta G} \right) \mathbf{w}_k\sqrt{a _{k,\widetilde{i}}}s_{k,\widetilde{i}}}}+\underset{\text{inter}-\text{cluster}~ \text{interference}}{\underbrace{\left( \mathbf{g}_{k,i}^{H}\mathbf{\Theta G} \right) \sum_{\widetilde{k}\ne k}^K{\mathbf{w}_{\widetilde{k}}\sum_{j\in \left\{ n,f \right\}}{\sqrt{a _{\widetilde{k},j}}s_{\widetilde{k},j}}}}}+ \underset{\text{noise}}{\underbrace{z_{k,i}}},
	\end{split}
\end{equation}
with $\widetilde{i},i\in \left\{ n,f \right\}$ and $\widetilde{i}\ne i$, where $\mathbf{\Theta }=\mathrm{diag}\left( \mathbf{v} \right) $ is the RIS's diagonal phase shift matrix and $z_{k,i}\sim \mathcal{C} \mathcal{N} \left( 0,\sigma ^2 \right) $ is the additive white Gaussian noise (AWGN).

In the proposed RIS-NOMA-ISAC system, SIC is applied for the users to decode their signals in each cluster. We assume a fixed decoding order in each cluster. Particularly, user $\mathrm{U}\left( k,n \right) $ first decodes the signal of user $\mathrm{U}\left( k,f \right) $ and then subtracts it from its observation to decode its own information. Therefore, the achievable rate for user $\mathrm{U}\left( k,n \right) $ to decode the signal of user $\mathrm{U}\left( k,f \right) $ is given by
\begin{equation}\label{Rn21} 
  \setlength{\abovedisplayskip}{5pt}
\setlength{\belowdisplayskip}{5pt}
	R_{k,f\rightarrow n}=\log _2\left( 1+\frac{a_{k,f}\left| \mathbf{g}_{k,n}^{H}\mathbf{\Theta Gw}_k \right|^2}{I_{k,n}^{\mathrm{itra}}+I_{k,n}^{\mathrm{iter}}+\sigma ^2} \right),  
\end{equation} 
where $I_{k,n}^{\mathrm{itra}}=a_{k,n}\left| \mathbf{g}_{k,n}^{H}\mathbf{\Theta Gw}_k \right|^2
$ and $I_{k,n}^{\mathrm{iter}}=\sum_{\widetilde{k}\ne k}^K{\left| \mathbf{g}_{k,n}^{H}\mathbf{\Theta Gw}_{\widetilde{k}} \right|^2}
$.

If the above decoding is successful, user $\mathrm{U}\left( k,n \right) $ removes the signal $s_{k,f}$ from $y_{k,n}$ to further decode its own signal $s_{k,n}$. Therefore, the individual achievable rate of user $\mathrm{U}\left( k,n \right) $ is given by
\begin{equation}\label{Rn1} 
  \setlength{\abovedisplayskip}{5pt}
\setlength{\belowdisplayskip}{5pt}
	R_{k,n}=\log _2\left( 1+\frac{a_{k,n}\left| \mathbf{g}_{k,n}^{H}\mathbf{\Theta Gw}_k \right|^2}{I_{k,n}^{\mathrm{iter}}+\sigma ^2} \right).    
\end{equation} 

Accordingly, user $\mathrm{U}\left( k,f \right) $ directly decodes its own signal by treating the other users' signals as interference. Therefore, the achievable rate for user $\mathrm{U}\left( k,f \right) $ to decode its own signal can be expressed as
\begin{equation}\label{Rn22}
  \setlength{\abovedisplayskip}{5pt}
 \setlength{\belowdisplayskip}{5pt}
	R_{k,f\rightarrow f}=\log _2\left( 1+\frac{a_{k,f}\left| \mathbf{g}_{k,f}^{H}\mathbf{\Theta Gw}_k \right|^2}{I_{k,f}^{\mathrm{itra}}+I_{k,f}^{\mathrm{iter}}+\sigma ^2} \right),  
\end{equation} 
where $I_{k,f}^{\mathrm{itra}}=a_{k,n}\left| \mathbf{g}_{k,f}^{H}\mathbf{\Theta Gw}_k \right|^2
$ and $I_{k,f}^{\mathrm{iter}}=\sum_{\widetilde{k}\ne k}^K{\left| \mathbf{g}_{k,f}^{H}\mathbf{\Theta Gw}_{\widetilde{k}} \right|^2}$.

As a result, the individual achievable rate of user $\mathrm{U}\left( k,f \right) $ is given by~\cite{9240028,8684765}
\begin{equation}\label{Rn2}
  \setlength{\abovedisplayskip}{5pt}
\setlength{\belowdisplayskip}{5pt}
	R_{k,f}=\min \left\{ R_{k,f\rightarrow n},R_{k,f\rightarrow f} \right\}.   
\end{equation}
	\vspace{-1.5cm}
\subsection{Radar Detection Model} 
 Since all the potential targets are in the NLoS \textcolor[rgb]{0.00,0.00,0.00}{areas} of the BS, no LoS links from the BS can be exploited to perform radar target sensing. However, we can utilize the reflection-LoS links created by the RIS to complete the sensing tasks. We consider using the RIS's beampattern gain as the sensing performance metric. The reflected signal at the RIS can be expressed as
 \begin{equation}\label{signal_RIS} 
  \setlength{\abovedisplayskip}{5pt}
\setlength{\belowdisplayskip}{5pt}
 	\overline{\mathbf{x}}=\mathbf{\Theta G}\left( \sum_{k=1}^K{\mathbf{w}_k\left( \sqrt{a _{k,n}}s_{k,n}+\sqrt{a _{k,f}}s_{k,f} \right)}  \right). 
 \end{equation}
 
 Therefore, the corresponding covariance matrix is given by
 \begin{equation}\label{covariance matrix}  
  \setlength{\abovedisplayskip}{5pt}
  \setlength{\belowdisplayskip}{5pt}
 	\mathbf{R}_{\overline{\mathbf{x}}}=\mathbb{E}\left( \overline{\mathbf{x}}\overline{\mathbf{x}}^H \right) =\mathbf{\Theta G}\left( \sum_{k=1}^K{\mathbf{w}_k\mathbf{w}_{k}^{H}}  \right) \mathbf{G}^H\mathbf{\Theta }^H.
 \end{equation}
 
 In our considered system, the communication signal is used to perform radar target sensing, thus the RIS's beampattern gain~\cite{Song2022JointTA} with respect
 to the $q$-th \textcolor[rgb]{0.00,0.00,0.00}{angle of interest}, i.e,  $\theta _{q}^{\text{Tg}}$, is given by
 \begin{equation}\label{beam pattern}  
  \setlength{\abovedisplayskip}{5pt}
  \setlength{\belowdisplayskip}{5pt}
 	\mathcal{P}_{\theta _{q}^{\text{Tg}}}\left( \mathbf{w}_k,\mathbf{v} \right) =\boldsymbol{a }^H\left( \theta _{q}^{\text{Tg}} \right) \mathbf{\Theta G}\left( \sum_{k=1}^K{\mathbf{w}_k\mathbf{w}_{k}^{H}}  \right) \mathbf{G}^H\mathbf{\Theta }^H\boldsymbol{a }\left( \theta _{q}^{\text{Tg}} \right), 
 \end{equation} 
 where $\boldsymbol{a }\left( \theta \right) =\left[ 1,e^{j\frac{2\pi d}{\lambda}\sin \left( \theta \right)},\cdots ,e^{j\frac{2\pi d}{\lambda}\left( M-1 \right) \sin \left( \theta \right)} \right] ^T$ is the steering vector at the RIS with angle $\theta$, $d$ denotes the antenna spacing, $\lambda $ denotes the carrier wavelength, $q\in \mathcal{Q} \triangleq \left\{ 1,2,\cdots ,Q \right\} 
 $, and $\mathcal{Q} _{\theta}=\left\{ \theta _{1}^{\mathrm{Tg}},\theta _{2}^{\mathrm{Tg}},\cdots ,\theta _{Q}^{\mathrm{Tg}} \right\} 
 $ is the set of sensing \textcolor[rgb]{0.00,0.00,0.00}{angles of interest}. 
 	\vspace{-0.5cm}
\subsection{Maximize the Minimum Beampattern Gain} 
The objective of this paper is to maximize the minimum beampattern gain towards \textit{Q} \textcolor[rgb]{0.00,0.00,0.00}{angles of interest} by jointly
optimizing the active beamforming, power allocation coefficients and passive beamforming.~Accordingly, the optimization problem is formulated as
\begin{subequations}\label{OP_MM}
  \setlength{\abovedisplayskip}{5pt}
  \setlength{\belowdisplayskip}{5pt}
	\begin{align}
		&\underset{\mathbf{a}_k,\mathbf{w}_k,\mathbf{v}}{\max}\underset{q\in \mathcal{Q}}{\min}\mathcal{P}_{\theta _{q}^{\text{Tg}}}\left( \mathbf{w}_k,\mathbf{v} \right) , \\
		&\mathrm{s}.\mathrm{t}.~R_{k,i}\geqslant R_{k,i}^{\min}, \label{OP_MM:b} \\
		&   \ \ \ \ \   \sum_{k=1}^K{\lVert \mathbf{w}_k \rVert _2}   \leqslant P_{\max}
		, \label{OP_MM:c}  \\
		&   \ \ \ \ \  a _{k,n}+a _{k,f}=1, a_{k,i}\in \left( 0,1 \right) , \label{OP_MM:d}  \\	
		&   \ \ \ \ \   \theta _{m}^{\text{RIS}}\in \left[ 0,2\pi \right) ,\label{OP_MM:e} 
	\end{align} 
\end{subequations}   
where $ R_{k,i}^{\min}$ is the minimum QoS requirement, $P_{\max}$ is the maximum transmit power at the BS, $\mathbf{a}_k=\left[ a_{k,n}~a_{k,f} \right] ^T$, $i\in \left\{ n,f \right\}, k\in \mathcal{K}$, $m\in \mathcal{M}$.
Constraint~\eqref{OP_MM:b} ensures the minimum QoS requirement of each NOMA user, constraint~\eqref{OP_MM:c} limits the maximum transmit power at the BS, constraint~\eqref{OP_MM:d} represents the power allocation coefficient constraint in each cluster, \textcolor[rgb]{0.00,0.00,0.00}{and} constraint~\eqref{OP_MM:e} is the
phase shift constraint for the RIS. 

Due to the coupling between $\mathbf{w}_k$, $\mathbf{a}_k$ and $\mathbf{v}$, the non-convexity of QoS constraints, and the exponential form of phase shifts in $\mathbf{v}$, problem~\eqref{OP_MM} is a highly nonconvex optimization problem and the optimal solution is in general intractable. In the next section, a suboptimal IBCD algorithm is proposed to solve problem~\eqref{OP_MM}.  
	\vspace{-0.5cm}
 \section{Proposed Solution}
 In this section, we elaborate on how to solve problem~\eqref{OP_MM}. We first transform problem~\eqref{OP_MM} into a more tractable form. To facilitate the design, we define $\mathbf{W}_k=\mathbf{w}_k\mathbf{w}_{k}^{H}$, where $\mathbf{W}_k$ satisfies $\mathbf{W}_k\succcurlyeq \textbf{0}$ and $\text{rank}\left( \mathbf{W}_k \right) =1$, $k\in \mathcal{K}$. Similarly, we define $\mathbf{V}=\mathbf{v}^H\mathbf{v}
 $, which needs to satisfy $\mathbf{V}\succcurlyeq \textbf{0}$ and $\text{rank}\left( \mathbf{V} \right) =1$. Then, the beampattern gain in~\eqref{beam pattern} can be rewritten as:
 \begin{equation}\label{beam pattern rewritten}  
    \setlength{\abovedisplayskip}{5pt}
    \setlength{\belowdisplayskip}{5pt}
 	\mathcal{P}_{\theta _{q}^{\text{Tg}}}\left( \mathbf{W}_k,\mathbf{V} \right) =\text{Tr}\left[ \mathbf{V\Upsilon }_q\left( \sum_{k=1}^K{\mathbf{W}_k}  \right) \mathbf{\Upsilon }_{q}^{H} \right] , 
 \end{equation} 
 where $\mathbf{\Upsilon }_q=\mathrm{diag}\left\{ \boldsymbol{a }^H\left( \theta _{q}^{\text{Tg}} \right) \right\} \mathbf{G}$.
 
 Moreover, the quadratic terms $\left| \mathbf{g}_{k,i}^{H}\mathbf{\Theta Gw}_k \right|^2
 $ and $\left| \mathbf{g}_{k,i}^{H}\mathbf{\Theta Gw}_{\widetilde{k}} \right|^2
 $ in~\eqref{Rn21},~\eqref{Rn1} and~\eqref{Rn22} can be respectively rewritten as:
 \begin{equation}\label{channel gain rewritten}
    \setlength{\abovedisplayskip}{5pt}
 	\setlength{\belowdisplayskip}{5pt}
 	\begin{cases}
 		\left| \mathbf{g}_{k,i}^{H}\mathbf{\Theta Gw}_k \right|^2=\text{Tr}\left( \mathbf{V\Gamma }_{k,i}\mathbf{W}_k\mathbf{\Gamma }_{k,i}^{H} \right),\\
 		\left| \mathbf{g}_{k,i}^{H}\mathbf{\Theta Gw}_{\widetilde{k}} \right|^2=\text{Tr}\left( \mathbf{V\Gamma }_{k,i}\mathbf{W}_{\widetilde{k}}\mathbf{\Gamma }_{k,i}^{H} \right), k\ne \tilde{k},\\
 	\end{cases} 
 \end{equation} 
 where $\mathbf{\Gamma }_{k,i}=\mathrm{diag}\left\{ \mathbf{g}_{k,i}^{H} \right\} \mathbf{G}$, $i\in \left\{ n,f \right\}$, and $k,\tilde{k} \in \mathcal{K}$.
 
 By exploiting the above definitions, constraints in~\eqref{OP_MM:b} can be equivalently reformulated as:
 \begin{equation}\label{R11 constrait} 
 	  \setlength{\abovedisplayskip}{5pt}
 	  \setlength{\belowdisplayskip}{5pt}
 	a_{k,n}\mathrm{Tr}\left( \mathbf{V\Gamma }_{k,n}\mathbf{W}_k\mathbf{\Gamma }_{k,n}^{H} \right) \geqslant r_{k,n}^{\min}\left( I_{k,n}^{\mathrm{iter}}+\sigma ^2 \right), 
 \end{equation}  
 \begin{equation}\label{R21 constrait} 
 	  \setlength{\abovedisplayskip}{5pt}
 	  \setlength{\belowdisplayskip}{5pt}
 	  a_{k,f}\mathrm{Tr}\left( \mathbf{V\Gamma }_{k,n}\mathbf{W}_k\mathbf{\Gamma }_{k,n}^{H} \right) \geqslant r_{k,f}^{\min}\left( I_{k,n}^{\mathrm{itra}}+I_{k,n}^{\mathrm{iter}}+\sigma ^2 \right), 
 \end{equation} 
 \begin{equation}\label{R22 constrait} 
 	  \setlength{\abovedisplayskip}{5pt}
 	   \setlength{\belowdisplayskip}{5pt}
 	a_{k,f}\mathrm{Tr}\left( \mathbf{V\Gamma }_{k,f}\mathbf{W}_k\mathbf{\Gamma }_{k,f}^{H} \right) \geqslant r_{k,f}^{\min}\left( I_{k,f}^{\mathrm{itra}}+I_{k,f}^{\mathrm{iter}}+\sigma ^2 \right),  
 \end{equation}   
 where $I_{k,n}^{\mathrm{itra}}=a_{k,n}\mathrm{Tr}\left( \mathbf{V\Gamma }_{k,n}\mathbf{W}_k\mathbf{\Gamma }_{k,n}^{H} \right) 
 $, $I_{k,f}^{\mathrm{itra}}=a_{k,n}\mathrm{Tr}\left( \mathbf{V\Gamma }_{k,f}\mathbf{W}_k\mathbf{\Gamma }_{k,f}^{H} \right) 
 $, $I_{k,i}^{\mathrm{iter}}=\sum_{\widetilde{k}\ne k}^K{\mathrm{Tr}\left( \mathbf{V\Gamma }_{k,i}\mathbf{W}_{\widetilde{k}}\mathbf{\Gamma }_{k,i}^{H} \right)}
 $, $r_{k,i}^{\min}=2^{R_{k,i}^{\min}}-1
 $.
 
Finally, the original problem~\eqref{OP_MM} can be equivalently recast as follows: 
\begin{subequations}\label{OP_MM_SDP}
  \setlength{\abovedisplayskip}{5pt}
  \setlength{\belowdisplayskip}{5pt}
	\begin{align}
		&\underset{\mathbf{a}_k,\mathbf{W}_k\succcurlyeq 0,\mathbf{V}\succcurlyeq 0}{\max}\underset{q\in \mathcal{Q}}{\min}\mathcal{P}_{\theta _{q}^{\text{Tg}}}\left( \mathbf{W}_k,\mathbf{V} \right) , \\
		&\mathrm{s}.\mathrm{t}.~\sum_{k=1}^K{\text{Tr}\left( \mathbf{W}_k \right)} \leqslant P_{\max}, \label{OP_MM_SDP:b} \\
		&   \ \ \ \ \  \left[ \mathbf{V} \right] _{m,m}=1, m\in \mathcal{M}, \label{OP_MM_SDP:c}\\
		&   \ \ \ \ \  \text{rank}\left( \mathbf{W}_k \right) =1, k\in \mathcal{K}, \label{OP_MM_SDP:d} \\
		&   \ \ \ \ \  \text{rank}\left( \mathbf{V} \right) =1, \label{OP_MM_SDP:e} 	\\
		&   \ \ \ \ \  \eqref{OP_MM:d},~\eqref{R11 constrait},~\eqref{R21 constrait},~\eqref{R22 constrait}. 
	\end{align} 
\end{subequations}  
 
 It is noted that the objective function of problem~\eqref{OP_MM_SDP} is non-smooth, \textcolor[rgb]{0.00,0.00,0.00}{hence,} we introduce an auxiliary variable $\chi$ to transform the original nonsmooth optimization problem~\eqref{OP_MM_SDP} into a smooth optimization problem, which is given by
 \begin{subequations}\label{OP_MM_SDP_smooth}
  \setlength{\abovedisplayskip}{5pt}
  \setlength{\belowdisplayskip}{5pt}
 	\begin{align}
 		&\underset{\mathbf{a }_k,\chi >0,\mathbf{W}_k\succcurlyeq 0,\mathbf{V}\succcurlyeq 0}{\max}\chi , \\
 		&\mathrm{s}.\mathrm{t}.~\text{Tr}\left[ \mathbf{V\Upsilon }_q\left( \sum_{k=1}^K{\mathbf{W}_k}  \right) \mathbf{\Upsilon }_{q}^{H} \right]  \geqslant \chi, q\in \mathcal{Q}, \label{OP_MM_SDP_smooth:b} \\ 		
 		&   \ \ \ \ \ \eqref{OP_MM:d},~\eqref{R11 constrait},~\eqref{R21 constrait},~\eqref{R22 constrait},~\eqref{OP_MM_SDP:b},~\eqref{OP_MM_SDP:c},~\eqref{OP_MM_SDP:d},~\eqref{OP_MM_SDP:e}. 
 	\end{align} 
 \end{subequations}  

 In the remainder of this article, we utilize \textcolor[rgb]{0.00,0.00,0.00}{the} IBCD approach to solve the equivalent problem~\eqref{OP_MM_SDP_smooth} instead of the original problem~\eqref{OP_MM}. To be specific, the coupled variables are divided into two blocks, namely $\left\{ \mathbf{W}_k,\mathbf{a}_k \right\} $ and $\mathbf{V}$. The active beamforming matrices $\left\{ \mathbf{W}_k \right\} $ and power allocation coefficients $\left\{ \mathbf{a}_k \right\} $ are first jointly optimized by applying \textcolor[rgb]{0.00,0.00,0.00}{the} SCA and SDR methods. Then, the passive beamforming matrix $\mathbf{V}$ is solved by employing the SRCR algorithm~\cite{Cao2017ASC}. 
\subsection{Joint Active Beamforming and Power Allocation Coefficients Optimization}
For a given passive beamforming matrix $\mathbf{V}$, the joint optimization problem for the design of $\left\{ \mathbf{W}_k \right\} $ and  $\left\{ \mathbf{a}_k \right\} $ is given by
\begin{subequations}\label{ABP_Max_SDP}
  \setlength{\abovedisplayskip}{5pt}
  \setlength{\belowdisplayskip}{5pt}
	\begin{align}
		&\underset{\mathbf{a }_k,\chi >0,\mathbf{W}_k\succcurlyeq 0}{\max}\chi , \\
		&\mathrm{s}.\mathrm{t}.~\eqref{OP_MM:d},~\eqref{R11 constrait},~\eqref{R21 constrait},~\eqref{R22 constrait},~\eqref{OP_MM_SDP:b},~\eqref{OP_MM_SDP:d},~\eqref{OP_MM_SDP_smooth:b}. 	 
	\end{align} 
\end{subequations}  

We note that the objective function of problem~\eqref{ABP_Max_SDP} is an affine function. However, $\mathbf{W}_k$ and $\mathbf{a}_k$ are coupled together in constraints~\eqref{R11 constrait},~\eqref{R21 constrait}, and~\eqref{R22 constrait}. Furthermore, the rank-one constraint~\eqref{OP_MM_SDP:d} is also nonconvex. As a result, these obstacles make problem~\eqref{ABP_Max_SDP} difficult to be solved. In the following, we will further transform and approximate problem~\eqref{ABP_Max_SDP} to achieve a tractable formulation.

To begin with, we define $\mathbf{H}_{k,i}=\mathbf{\Gamma }_{k,i}^{H}\mathbf{V\Gamma }_{k,i}$, $i\in \left\{ n,f \right\} $. In addition, according to the equality constraint~\eqref{OP_MM:d}, we have: $a _{k,f}=1-a _{k,n}$. Then, substituting $\mathbf{H}_{k,i}$ and $a _{k,f}$ into~\eqref{R11 constrait},~\eqref{R21 constrait} and~\eqref{R22 constrait}, we have the following equivalent inequality constraints,
\begin{equation}\label{R11_equivalent}
    \setlength{\abovedisplayskip}{5pt}
	\setlength{\belowdisplayskip}{5pt}
	a_{k,n}\mathrm{Tr}\left( \mathbf{W}_k\mathbf{H}_{k,n} \right) 
	\geqslant r_{k,n}^{\min}\left( I_{k,n}^{\mathrm{iter}}+\sigma ^2 \right), 
\end{equation} 
\begin{equation}\label{R21_equivalent}
    \setlength{\abovedisplayskip}{5pt}
	\setlength{\belowdisplayskip}{5pt}
	\frac{r_{k,f}^{\min}\left( \frac{\mathrm{Tr}\left( \mathbf{W}_k\mathbf{H}_{k,n} \right)}{r_{k,f}^{\min}}-I_{k,n}^{\mathrm{iter}}-\sigma ^2 \right)}{r_{k,f}^{\min}+1}\geqslant a_{k,n}\mathrm{Tr}\left( \mathbf{W}_k\mathbf{H}_{k,n} \right) ,
\end{equation} 
\begin{equation}\label{R22_equivalent}
	\setlength{\abovedisplayskip}{5pt}
	\setlength{\belowdisplayskip}{5pt}
	\frac{r_{k,f}^{\min}\left( \frac{\mathrm{Tr}\left( \mathbf{W}_k\mathbf{H}_{k,f} \right)}{r_{k,f}^{\min}}-I_{k,f}^{\mathrm{iter}}-\sigma ^2 \right)}{r_{k,f}^{\min}+1}\geqslant a_{k,n}\mathrm{Tr}\left( \mathbf{W}_k\mathbf{H}_{k,f} \right),
\end{equation} 
where $I_{k,i}^{\mathrm{iter}}=\sum_{\widetilde{k}\ne k}^K{\mathrm{Tr}\left( \mathbf{W}_{\widetilde{k}}\mathbf{H}_{k,i} \right)}$.

It is observed that the functions in the left hand side of~\eqref{R11_equivalent}, and in the right hand sides of~\eqref{R21_equivalent} and~\eqref{R21_equivalent} have similar structure. In fact, the functions $a_{k,n}\mathrm{Tr}\left( \mathbf{W}_k\mathbf{H}_{k,n} \right) $ and $a_{k,n}\mathrm{Tr}\left( \mathbf{W}_k\mathbf{H}_{k,f} \right)$ are bilinear functions over $a_{k,n}$ and $\mathbf{W}_k$, which are neither convex nor concave. To develop \textcolor[rgb]{0.00,0.00,0.00}{an} efficient algorithm for jointly optimizing $\left\{ \mathbf{W}_k \right\}$ and $\left\{ \mathbf{a}_k \right\}$, \textcolor[rgb]{0.00,0.00,0.00}{the} SCA and SDR approaches are leveraged to tackle the nonconvex constraints~\eqref{R11_equivalent}, ~\eqref{R21_equivalent}, and ~\eqref{R22_equivalent}. In particular, we first reformulate constraint~\eqref{R11_equivalent} by introducing a new slack variable $\eta _k$, such that
\begin{numcases}{}
	  \setlength{\abovedisplayskip}{5pt}
	\setlength{\belowdisplayskip}{5pt}
	a_{k,n}\mathrm{Tr}\left( \mathbf{W}_k\mathbf{H}_{k,n} \right) \geqslant \eta _{k}^{2}, \label{slack_1}\\ 
	\eta _{k}^{2}\geqslant r_{k,n}^{\min}\left( I_{k,n}^{\mathrm{iter}}+\sigma ^2 \right). \label{slack_2}
\end{numcases}

Then, by applying the Schur complement theory~\cite{Xie2021JointOO}, \eqref{slack_1} can be
further expressed in \textcolor[rgb]{0.00,0.00,0.00}{the} linear matrix inequality form, which is given by
\begin{equation}\label{LMI}
	\left[ \begin{matrix}
		a_{k,n}&		\eta _k\\
		\eta _k&		\mathrm{Tr}\left( \mathbf{W}_k\mathbf{H}_{k,n} \right)\\
	\end{matrix} \right] \succcurlyeq \textbf{0},\exists \eta _k> 0.
\end{equation} 

Furthermore, by using the SCA approach based on \textcolor[rgb]{0.00,0.00,0.00}{first}-order Taylor approximation, \eqref{slack_2} can be approximated as follows
\begin{equation}\label{AB_Max_SDP:c21}
	  \setlength{\abovedisplayskip}{5pt}
	\setlength{\belowdisplayskip}{5pt}
	\widetilde{\eta _{k}^{2}}+2\widetilde{\eta _k}\left( \eta _k-\widetilde{\eta _k} \right) \geqslant r_{k,n}^{\min}\left( I_{k,n}^{\mathrm{iter}}+\sigma ^2 \right),
\end{equation} 
where $\widetilde{\eta }_{k}$ is a fixed point and can be updated by $\widetilde{\eta }_{k}^{\left( t_1 \right)}=\eta _{k}^{\left( t_1 \right)}$, \textcolor[rgb]{0.00,0.00,0.00}{where} $t_1$ is the iteration index. 

Next, we tackle the non-convexity in constraints~\eqref{R21_equivalent} and~\eqref{R22_equivalent}. To tackle this challenge, we resort to
the arithmetic-geometric mean  inequality~\cite{Sun2018JointBA}. Specifically, $a _{k,n}\text{Tr}\left( \mathbf{W}_k\mathbf{H}_{k,n} \right) 
$ and $a _{k,n}\text{Tr}\left( \mathbf{W}_k\mathbf{H}_{k,f} \right)$ can be\textcolor[rgb]{0.00,0.00,0.00}{, respectively,} approximated as 
\begin{equation}\label{CUB1}
  \setlength{\abovedisplayskip}{5pt}
\setlength{\belowdisplayskip}{5pt}
	a_{k,n}\mathrm{Tr}\left( \mathbf{W}_k\mathbf{H}_{k,n} \right) \leqslant \frac{\beta _{k,1}a_{k,n}^{2}}{2}+\frac{\left( \mathrm{Tr}\left( \mathbf{W}_k\mathbf{H}_{k,n} \right) \right) ^2}{2\beta _{k,1}}\triangleq \mathcal{T} _{k,1},
\end{equation} 
\begin{equation}\label{CUB2}
	a_{k,n}\mathrm{Tr}\left( \mathbf{W}_k\mathbf{H}_{k,f} \right) \leqslant \frac{\beta _{k,2}a_{k,n}^{2}}{2}+\frac{\left( \mathrm{Tr}\left( \mathbf{W}_k\mathbf{H}_{k,f} \right) \right) ^2}{2\beta _{k,2}}\triangleq \mathcal{T} _{k,2},
\end{equation} 
where $\beta _{k,1}$ and $\beta _{k,2}$ are fixed points. The equality in~\eqref{CUB1} and~\eqref{CUB2} will always hold if $\beta _{k,1}=\frac{\text{Tr}\left( \mathbf{W}_k\mathbf{H}_{k,n} \right)}{a _{k,n}}$ and $\beta _{k,2}=\frac{\text{Tr}\left( \mathbf{W}_k\mathbf{H}_{k,f} \right)}{a _{k,n}}
$.

Based on the aforementioned transformations and approximations, \textcolor[rgb]{0.00,0.00,0.00}{the} constraints given in~\eqref{R21_equivalent} and~\eqref{R22_equivalent} can be\textcolor[rgb]{0.00,0.00,0.00}{, respectively,} reformulated as
\begin{equation}\label{AB_Max_SDP:d11}
  \setlength{\abovedisplayskip}{0pt}
\setlength{\belowdisplayskip}{0pt}
	\frac{r_{k,f}^{\min}\left( \frac{\mathrm{Tr}\left( \mathbf{W}_k\mathbf{H}_{k,n} \right)}{r_{k,f}^{\min}}-I_{k,n}^{\mathrm{iter}}-\sigma ^2 \right)}{r_{k,f}^{\min}+1}\geqslant \mathcal{T} _{k,1},
\end{equation} 
\begin{equation}\label{AB_Max_SDP:e11}
  \setlength{\abovedisplayskip}{0pt}
   \setlength{\belowdisplayskip}{0pt}
	\frac{r_{k,f}^{\min}\left( \frac{\mathrm{Tr}\left( \mathbf{W}_k\mathbf{H}_{k,f} \right)}{r_{k,f}^{\min}}-I_{k,f}^{\mathrm{iter}}-\sigma ^2 \right)}{r_{k,f}^{\min}+1}\geqslant \mathcal{T} _{k,2}.
\end{equation} 

Finally, let us turn our attention to the rank-one constraint~\eqref{OP_MM_SDP:d}.  
To address this issue, we exploit the SDR technique by removing the rank-one constraints from the problem formulation. In particular, the relaxed problem is \textcolor[rgb]{0.00,0.00,0.00}{given} as
\begin{subequations}\label{AB_Max_SDR}
  \setlength{\abovedisplayskip}{5pt}
\setlength{\belowdisplayskip}{5pt}
	\begin{align}
		&\underset{\chi ,\eta _k>0,0<a_{k,n}<1,\mathbf{W}_k\succcurlyeq 0}{\max}\chi, \\
		&\mathrm{s}.\mathrm{t}.~~\eqref{OP_MM_SDP:b},~\eqref{OP_MM_SDP_smooth:b},~\eqref{LMI},~\eqref{AB_Max_SDP:c21}, ~\eqref{AB_Max_SDP:d11},~\eqref{AB_Max_SDP:e11}. 
	\end{align} 
\end{subequations}   

Obviously, problem~\eqref{AB_Max_SDR} is a convex SDP problem and can be efficiently solved by the CVX tool~\cite{cvx}. In the following theorem, we will verify the tightness of the relaxed problem~\eqref{AB_Max_SDR}.
\begin{theorem}\label{Rank1 in joint problem}
	If the relaxed problem~\eqref{AB_Max_SDR} is feasible, then the solutions $\left\{ \mathbf{W}_k \right\} $ obtained by solving problem~\eqref{AB_Max_SDR} always satisfy $\text{rank}\left( \mathbf{W}_k \right) \leqslant 1$, $k\in \mathcal{K}$.
	
	\textit{Proof}: Please refer to Appendix A. 
\end{theorem}
	\vspace{-0.3cm}
	
\textbf{Theorem~\ref{Rank1 in joint problem}} represents the fact that we can obtain the rank-one solutions of problem~\eqref{ABP_Max_SDP} by solving the convex problem~\eqref{AB_Max_SDR}.
	\vspace{-0.3cm}
\begin{algorithm}
	\caption{Proposed joint active beamforming and power allocation coefficients optimization algorithm}
	\label{ABPAC}
	\begin{algorithmic}[1]
		\STATE  \textbf{Initialize} $\beta _{k,1}^{\left( 0 \right)},\beta _{k,2}^{\left( 0 \right)},\widetilde{\eta }_{k}^{\left( 0 \right)},k\in \mathcal{K}
		$, and set ${t_1} = 0$.
		\REPEAT
		\STATE   ${t_1} = {t_1} + 1$;
		\STATE   update $ \mathbf{W}_{k}^{\left( t_1 \right)}$, $a_{k,n}^{\left( t_1 \right)}$ and $  \eta _{k}^{\left( t_1 \right)}$ by solving problem~\eqref{AB_Max_SDR} with given $\beta _{k,1}^{\left( t_1-1 \right)},\beta _{k,2}^{\left( t_1-1 \right)},\widetilde{\eta }_{k}^{\left( t_1-1 \right)}$;
		\STATE  update  $\widetilde{\eta }_{k}^{\left( t_1 \right)}=\eta _{k}^{\left( t_1 \right)}$, $\beta _{k,1}^{\left( t_1 \right)}=\frac{\text{Tr}\left( \mathbf{W}_{k}^{\left( t_1 \right)}\mathbf{H}_{k,n} \right)}{a _{k,n}^{\left( t_1 \right)}}$ and $\beta _{k,2}^{\left( t_1 \right)}=\frac{\text{Tr}\left( \mathbf{W}_{k}^{\left( t_1 \right)}\mathbf{H}_{k,f} \right)}{a _{k,n}^{\left( t_1 \right)}} $;
		\UNTIL {the objective value of problem~\eqref{AB_Max_SDR} converge}.
		\STATE  \textbf{Output}: $\mathbf{W}_{k} $ and $\mathbf{a}_{k}$, $ k\in \mathcal{K}$.
	\end{algorithmic}
\end{algorithm}
	\vspace{-0.5cm}
	
Based on the above discussion, the proposed algorithm to solve problem~\eqref{ABP_Max_SDP} is summarized in \textbf{Algorithm~\ref{ABPAC}}. It is noted that
we need to initialize the fixed points $\left\{ \beta _{k,1}^{\left( 0 \right)},\beta _{k,2}^{\left( 0 \right)},\widetilde{\eta }_{k}^{\left( 0 \right)} \right\} $ in \textbf{Algorithm~\ref{ABPAC}}. However, it is difficult to find these feasible fixed points. In the following, we proceed to construct a feasibility problem and develop a novel feasible initial points finding algorithm. By \textcolor[rgb]{0.00,0.00,0.00}{introducing} an infeasibility indicator $\delta \ge 0$, the formulated feasibility problem is given as:
\begin{subequations}\label{feasible problem}
	  \setlength{\abovedisplayskip}{5pt}
	\setlength{\belowdisplayskip}{5pt}
	\begin{align}
		&\underset{\delta \geqslant 0,\chi >0,0<a_{k,n}<1,\mathbf{W}_k\succcurlyeq \textbf{0}}{\min}\delta,\\
		&{\mathrm{s}.\mathrm{t}.}~\mathrm{Tr}\left[ \mathbf{V\Upsilon }_q\left( \sum_{k=1}^K{\mathbf{W}_k} \right) \mathbf{\Upsilon }_{q}^{H} \right] +\delta \geqslant \chi , \\
		&   \ \ \ \ \ \widetilde{\eta _{k}^{2}}+2\widetilde{\eta _k}\left( \eta _k-\widetilde{\eta _k} \right) +\delta \geqslant r_{k,n}^{\min}\left( I_{k,n}^{\mathrm{iter}}+\sigma ^2 \right) 
		,\\		
		& \ \ \ \ \   \frac{r_{k,f}^{\min}\left( \frac{\mathrm{Tr}\left( \mathbf{W}_k\mathbf{H}_{k,n} \right)}{r_{k,f}^{\min}}-I_{k,n}^{\mathrm{iter}}-\sigma ^2 \right)}{r_{k,f}^{\min}+1} +\delta \geqslant \mathcal{T} _{k,1}, \\
		& \ \ \ \ \ \frac{r_{k,f}^{\min}\left( \frac{\mathrm{Tr}\left( \mathbf{W}_k\mathbf{H}_{k,f} \right)}{r_{k,f}^{\min}}-I_{k,f}^{\mathrm{iter}}-\sigma ^2 \right)}{r_{k,f}^{\min}+1} +\delta \geqslant \mathcal{T} _{k,2},\\
		& \ \ \ \ \ P_{\max}+ \delta \geqslant \sum_{K=1}^K{\text{Tr}\left( \mathbf{W}_k \right)},\\ 	
		& \ \ \ \ \	\eqref{LMI},			
	\end{align}
\end{subequations}
where $\delta $ denotes how far the corresponding \textcolor[rgb]{0.00,0.00,0.00}{constraint} in~problem~\eqref{AB_Max_SDR} is from being satisfied, $q\in \mathcal{Q}$ and $k\in \mathcal{K}$.

Problem~\eqref{feasible problem} is also a convex optimization problem, which can be solved efficiently. The proposed feasible points finding algorithm is summarized in\textbf{ Algorithm 2}.
\vspace{-0.3cm}
\begin{algorithm}
	\caption{Feasible initial points finding algorithm}
	\label{Feasible_algorithm}
	\begin{algorithmic}[1]
		\STATE  Randomly initialize fixed points $\beta _{k,1}^{\left( 0 \right)},\beta _{k,2}^{\left( 0 \right)},\widetilde{\eta }_{k}^{\left( 0 \right)},k\in \mathcal{K}		$. Let iteration index ${t_2} = 0$.
		\REPEAT
		\STATE   ${t_2} = {t_2} + 1$;
		\STATE   update $ \mathbf{W}_{k}^{\left( t_2 \right)}$, $\mathbf{R}_{0}^{\left( t_2 \right)} $ and $  \eta _{k}^{\left( t_2 \right)}$ by solving problem~\eqref{feasible problem} with given $\beta _{k,1}^{\left( t_2-1 \right)},\beta _{k,2}^{\left( t_2-1 \right)},\widetilde{\eta }_{k}^{\left( t_2-1 \right)},k\in \mathcal{K}$;
		\STATE  update  $\widetilde{\eta }_{k}^{\left( t_2 \right)}=\eta _{k}^{\left( t_2 \right)}$, $\beta _{k,2}^{\left( t_2 \right)}=\frac{\text{Tr}\left( \mathbf{W}_{k}^{\left( t_2 \right)}\mathbf{H}_{k,n} \right)}{a _{k,n}^{\left( t_2 \right)}}$ and $\beta _{k,2}^{\left( t_2 \right)}=\frac{\text{Tr}\left( \mathbf{W}_{k}^{\left( t_2 \right)}\mathbf{H}_{k,f} \right)}{a _{k,n}^{\left( t_2 \right)}} $;
		\UNTIL {$\delta$ below a predefined threshold}.
		\STATE  \textbf{Output}: $\beta _{k,1},\beta _{k,2},\widetilde{\eta }_{k},k\in \mathcal{K}		$.
	\end{algorithmic}
\end{algorithm}
\begin{remark}\label{remark:algorithm 4}
	When $\delta\rightarrow 0$, the obtained solutions of problem~\eqref{feasible problem} are feasible for problem~\eqref{AB_Max_SDR}. Therefore, the output of \textbf{Algorithm~\ref{Feasible_algorithm}} can be used to replace the initial fixed points in \textbf{Algorithm~\ref{ABPAC}}.
\end{remark}
\subsection{Passive Beamforming Optimization} 
For any given  $\left\{ \mathbf{W}_k \right\}$ and $\left\{ \textbf{a} _{k} \right\}$, the passive beamforming optimization problem is given by
\begin{subequations}\label{PB_Max_SDP}
  \setlength{\abovedisplayskip}{5pt}
\setlength{\belowdisplayskip}{5pt}
	\begin{align}
		&\underset{\chi >0,\mathbf{V}\succcurlyeq \textbf{0}}{\max}\chi , \\
		&\mathrm{s}.\mathrm{t}.~\eqref{R11 constrait},~\eqref{R21 constrait},~\eqref{R22 constrait},~\eqref{OP_MM_SDP:c},~\eqref{OP_MM_SDP:e},~\eqref{OP_MM_SDP_smooth:b}.	
	\end{align} 
\end{subequations} 

Now, the remaining non-convexity in problem~\eqref{PB_Max_SDP} lies in
the rank-one constraint~\eqref{OP_MM_SDP:e}. According to the SRCR algorithm~\cite{Cao2017ASC}, the constraint $\text{rank}\left( \mathbf{V} \right) =1
$ can be transformed equivalently as:
\begin{equation} \label{rank_ralex} 
  \setlength{\abovedisplayskip}{5pt}
\setlength{\belowdisplayskip}{5pt}
	\mathbf{e}_{\max}^{H}\left( \mathbf{V}^{\left( t_3 \right)} \right) \mathbf{Ve}_{\max}\left( \mathbf{V}^{\left( t_3 \right)} \right) \geqslant \varepsilon ^{\left( t_3 \right)}\mathrm{Tr}\left( \mathbf{V} \right), 
\end{equation}   
where $\mathbf{V}^{\left( t_3 \right)}$ is the obtained solution in the $t_3$-th iteration, $\mathbf{e}_{\max}\left( \mathbf{V}^{\left( t_3 \right)} \right) $ is the eigenvector corresponding to the maximum eigenvalue of $\mathbf{V}^{\left( t _3 \right)}$, $\varepsilon ^{\left( t_3 \right)}\in \left[ 0,1 \right]$ is a relaxation parameter in the $t_3$-th iteration. We can increase $\varepsilon ^{\left( t_3 \right)}$
from 0 to 1 sequentially via iterations to gradually
approach a rank-one solution. After each iteration, the relaxation parameter can be updated as
\begin{equation} \label{update relaxation parameter} 
  \setlength{\abovedisplayskip}{5pt}
\setlength{\belowdisplayskip}{5pt}
	\varepsilon ^{\left( t_3+1 \right)}\longleftarrow \min \left( 1,\frac{\lambda _{\max}\left( \mathbf{V}^{\left( t_3+1 \right)} \right)}{\mathrm{Tr}\left( \mathbf{V}^{\left( t_3+1 \right)} \right)}+\rho ^{\left( t_3+1 \right)} \right) ,
\end{equation}  
where $\lambda _{\max}\left( \mathbf{V}^{\left( t_3 \right)} \right) $ is the largest eigenvalue of $\mathbf{V}^{\left( t_3 \right)}$ and $\rho ^{\left( t_3 \right)}$ denotes the step size.

Finally, in the $t_3$-th iteration, the optimization problem that needs to be solved is given as follows
\begin{subequations}\label{PB_Max_SROCR}
  \setlength{\abovedisplayskip}{5pt}
\setlength{\belowdisplayskip}{5pt}
	\begin{align}
		&\underset{\chi >0,\mathbf{V}\succcurlyeq \textbf{0}}{\max}\chi , \\
		&\mathrm{s}.\mathrm{t}.~\eqref{R11 constrait},~\eqref{R21 constrait},~\eqref{R22 constrait},~\eqref{OP_MM_SDP:c},~\eqref{OP_MM_SDP_smooth:b},~\eqref{rank_ralex} .  	
	\end{align} 
\end{subequations}  

Problem~\eqref{PB_Max_SROCR} is a SDP problem and can be solved by the CVX tool~\cite{cvx}. The procedure for optimizing passive beamforming is sketched in \textbf{Algorithm~\ref{PB_algorithm}}.
\begin{algorithm}  
	\caption{Proposed passive beamforming optimization algorithm}
	\label{PB_algorithm}
	\begin{algorithmic}[1]
		\STATE  Initialize $\mathbf{V}^{\left( 0 \right)}
		$ and $\rho ^{\left( 0 \right)}$. Set $\varepsilon ^{\left( t _3 \right)}=0$ and  ${t_3} = 0$.
		\REPEAT
		\STATE  Solve problem~\eqref{PB_Max_SROCR} with $\left\{ \varepsilon ^{\left( t_3 \right)},\mathbf{V}^{\left( t_3 \right)} \right\} 
		$ to obtain $\mathbf{V}^*$;
		\STATE \textbf{if} problem~\eqref{PB_Max_SROCR} is solvable 
		\STATE ~~~Update $\mathbf{V}^{\left( t_3+1 \right)}=\mathbf{V}^*$; 
		\STATE ~~~Update $\rho ^{\left( t_3+1 \right)}=\rho ^{\left( 0 \right)}
		$;
		\STATE \textbf{else}  
		\STATE ~~~Update $\mathbf{V}^{\left( t_3+1 \right)}=\mathbf{V}^{\left( t_3 \right)}
		$;  
		\STATE ~~~update $\rho ^{\left( t_3+1 \right)}=\frac{\rho ^{\left( t_3 \right)}}{2}
		$;
		\STATE \textbf{end}
		\STATE  Update $\varepsilon ^{\left( t _3 + 1 \right)}$ via~\eqref{update relaxation parameter};
		\STATE  Update ${t_3} = {t_3} + 1$;				
		\UNTIL {$\frac{\mathrm{Tr}\left( \mathbf{V}^{\left( t_3 \right)} \right)}{\lambda _{\max}\left( \mathbf{V}^{\left( t_3 \right)} \right)}$ is below a predefined threshold and the objective value of problem~\eqref{PB_Max_SROCR} converges.}
		\STATE   \textbf{Output}: $\mathbf{V}$.
	\end{algorithmic}
\end{algorithm} 	
\vspace{-0.3cm}
\subsection{Proposed Algorithm, Complexity and Convergence}
To facilitate the understanding of the proposed IBCD algorithm
for solving problem~\eqref{OP_MM_SDP_smooth}, we summarize it in~\textbf{Algorithm~\ref{proposed BCD algorithm}}. The convergence of \textbf{Algorithm~\ref{proposed BCD algorithm}} is analyzed as follows.~\textbf{Algorithm~\ref{ABPAC}} and~\textbf{Algorithm~\ref{PB_algorithm}} converge to a KKT stationary solution of problem~\eqref{ABP_Max_SDP} and problem~\eqref{PB_Max_SDP}, respectively. Similar \textcolor[rgb]{0.00,0.00,0.00}{proofs} can be found in~\cite{Sun2018JointBA} and~\cite{Cao2017ASC}.
Furthermore, the objective value of problem~\eqref{OP_MM_SDP_smooth} is non-decreasing after each iteration and the radar beampattern gain is upper bounded. Therefore, the proposed IBCD algorithm is guaranteed to converge. Let $T_{1}^{\max}$ and $T_{2}^{\max}$ denote the \textcolor[rgb]{0.00,0.00,0.00}{number of iterations} of~\textbf{Algorithm~\ref{ABPAC}} and~\textbf{Algorithm~\ref{Feasible_algorithm}}.
Then, the complexity of~\textbf{Algorithm~\ref{ABPAC}} is given by $\mathcal{O} \left( T_{1}^{\max}O_1+T_{2}^{\max}O_2 \right) $, where $O_i$ is defined as
\begin{equation} \label{O_i} 
	  \setlength{\abovedisplayskip}{5pt}
	\setlength{\belowdisplayskip}{5pt}
	O_i=\max \left\{ 3K+Q+1,N \right\} ^4\sqrt{N}\log \frac{1}{\varpi _i},
\end{equation}  
where $\varpi _1$ and $\varpi _2$ are the solution \textcolor[rgb]{0.00,0.00,0.00}{accuracies} of~\textbf{Algorithm~\ref{ABPAC}} and~\textbf{Algorithm~\ref{Feasible_algorithm}}, $i\in \left\{ 1,2 \right\} $. 

Similarly, the complexity of \textbf{Algorithm~\ref{PB_algorithm}} is given by $\mathcal{O} \left( T_{3}^{\max}O_3 \right) $ with
\begin{equation} \label{O_3} 
	  \setlength{\abovedisplayskip}{5pt}
	\setlength{\belowdisplayskip}{5pt}
  O_3=\max \left\{ 3K+Q+1,M \right\} ^4\sqrt{M}\log \frac{1}{\varpi _3},
\end{equation} 
where $T_{3}^{\max}$ and $\varpi _3$ are the corresponding \textcolor[rgb]{0.00,0.00,0.00}{number of iterations} and solution \textcolor[rgb]{0.00,0.00,0.00}{accuracy}, respectively. 

Finally, the total complexity of \textbf{Algorithm~\ref{proposed BCD algorithm}} is given by
\begin{equation} \label{O_4} 
	  \setlength{\abovedisplayskip}{5pt}
	\setlength{\belowdisplayskip}{5pt}
 \mathcal{O} \left( T_{4}^{\max}\left( T_{1}^{\max}O_1+T_{2}^{\max}O_2+T_{3}^{\max}O_3 \right) \right), 
\end{equation} 
 where $T_{4}^{\max}$ is the number of iterations of \textbf{Algorithm~\ref{proposed BCD algorithm}}.
\begin{algorithm}
	\caption{Proposed IBCD algorithm for solving problem~\eqref{OP_MM}}
	\label{proposed BCD algorithm}
	\begin{algorithmic}[1]
		\STATE  Initialize $\mathbf{V}^{\left( 0 \right)}$ and set ${t_4} = 0$.
		\REPEAT
		\STATE  ${t_4} = {t_4} + 1$;
		\STATE  update $\mathbf{W}_{k}^{\left( t_4 \right)} $ and $\mathbf{a}_{k}^{\left( t_4\right)}$ by \textbf{Algorithm~\ref{ABPAC}} with $\mathbf{V}^{\left( t_4-1 \right)}$; 
		\STATE  update $\mathbf{V}^{\left( {t_4 } \right)}$ by \textbf{Algorithm~\ref{PB_algorithm}} with $\mathbf{W}_{k}^{\left( t_4 \right)} $ and $\mathbf{a}_{k}^{\left( t_4\right)}$;
		\UNTIL {the objective value of problem~\eqref{OP_MM_SDP_smooth} converges.}
		\STATE   \textbf{Output}: $\mathbf{W}_{k} $, $\mathbf{a}_{k}$, and $\mathbf{V}$, $ k\in \mathcal{K}$.
	\end{algorithmic}
\end{algorithm} 
\vspace{-0.5cm}

 \section{Low Complexity Algorithm} 
 In the developed~\textbf{Algorithm~\ref{proposed BCD algorithm}}, $\left\{ \mathbf{W}_k \right\}  $ and $\left\{ \mathbf{a}_{k} \right\} $ are jointly optimized by applying~\textbf{Algorithm~\ref{ABPAC}}. However, the computational complexity of~\textbf{Algorithm~\ref{ABPAC}} scales \textcolor[rgb]{0.00,0.00,0.00}{linearly} with $T_{1}^{\max}$ and $T_{2}^{\max}$. Note that the joint optimization strategy over $\left\{ \mathbf{W}_k \right\}  $ and $\left\{ \mathbf{a}_{k} \right\} $  can achieve better performance than other benchmark schemes, which will be revealed via simulations in Section V. To strike a balance between the performance and computational complexity, we develop a low-complexity algorithm based on \textcolor[rgb]{0.00,0.00,0.00}{the} IAO method, where the active beamforming, power allocation coefficients and passive beamforming are optimized iteratively in an alternating manner. In each iteration, the passive beamforming solution is also obtained by~\textbf{Algorithm~\ref{PB_algorithm}}. \textcolor[rgb]{0.00,0.00,0.00}{However}, the active beamforming is optimized by a single SDP problem and the solutions of the power allocation coefficients are obtained by a proposed closed-form power allocation strategy. To begin with, we first exploit the feasibility of problem~\eqref{OP_MM_SDP_smooth}. 
\begin{theorem}\label{feaible_check_theorem}
	Given the active beamforming matrix $\left\{ \mathbf{W}_{k} \right\} $ and passive beamforming matrix $\mathbf{V}$, problem~\eqref{OP_MM_SDP_smooth} is feasible, if and only if the following inequality holds:
	\begin{equation}\label{feasible}
		  \setlength{\abovedisplayskip}{5pt}
		\setlength{\belowdisplayskip}{5pt}
		\max \left( a_{k,f}^{\min ,1},a_{k,f}^{\min ,2} \right)  \leqslant a_{k,f}^{\max}<1,
	\end{equation}
	where $a_{k,f}^{\max}$, $a_{k,f}^{\min ,1}$ and $a_{k,f}^{\min ,2}$ are\textcolor[rgb]{0.00,0.00,0.00}{, respectively,} defined as
	\begin{equation}\label{a_kf_max}
		  \setlength{\abovedisplayskip}{5pt}
		\setlength{\belowdisplayskip}{5pt}
		a_{k,f}^{\max}=1-\frac{r_{k,n}^{\min}\left[ \sum_{\widetilde{k}\ne k}^K{\mathrm{Tr}\left( \mathbf{W}_{\widetilde{k}}\mathbf{H}_{k,n} \right)}+\sigma ^2 \right]}{\mathrm{Tr}\left( \mathbf{W}_k\mathbf{H}_{k,n} \right)},
	\end{equation}
	\begin{equation}\label{a_kf_min1}
		  \setlength{\abovedisplayskip}{5pt}
		\setlength{\belowdisplayskip}{5pt}
		a_{k,f}^{\min ,1}=\frac{\mathrm{Tr}\left( \mathbf{W}_k\mathbf{H}_{k,n} \right) +\sum_{\widetilde{k}\ne k}^K{\left( \mathbf{W}_{\widetilde{k}}\mathbf{H}_{k,n} \right)}+\sigma ^2}{\left( 1+\frac{1}{r_{k,f}^{\min}} \right) \mathrm{Tr}\left( \mathbf{W}_k\mathbf{H}_{k,n} \right)}, 
	\end{equation}	
	\begin{equation}\label{a_kf_min2}
		  \setlength{\abovedisplayskip}{5pt}
		\setlength{\belowdisplayskip}{5pt}
		a_{k,f}^{\min ,2}=\frac{\mathrm{Tr}\left( \mathbf{W}_k\mathbf{H}_{k,f} \right) +\sum_{\widetilde{k}\ne k}^K{\mathrm{Tr}\left( \mathbf{W}_{\widetilde{k}}\mathbf{H}_{k,f} \right)}+\sigma ^2}{\left( 1+\frac{1}{r_{k,f}^{\min}} \right) \mathrm{Tr}\left( \mathbf{W}_k\mathbf{H}_{k,f} \right)}.
	\end{equation}
	
	\textit{Proof}: Please refer to Appendix~B.
\end{theorem}
	Following the idea introduced in~\cite{Zhu2019JointTB}, in each cluster, the power allocation should be first performed for the RFU and only necessary power should be allocated to satisfy the minimum QoS constraints. Then, the remaining power should be allocated to the RNU. According to the minimum QoS constraints \eqref{R21 constrait} and \eqref{R22 constrait}, if $a_{k,f}$ is a feasible solution of problem~\eqref{OP_MM_SDP_smooth}, we have: $a_{k,f}\geqslant \max \left\{ a_{k,f}^{\min ,1},a_{k,f}^{\min ,2} \right\} 
	$. Finally, the closed-form power allocation coefficients \textcolor[rgb]{0.00,0.00,0.00}{are} given by: 
	\begin{equation}\label{power_allocation_sub}
		\setlength{\abovedisplayskip}{5pt}
		\setlength{\belowdisplayskip}{5pt}
		a_{k,f}=\max \left\{ a_{k,f}^{\min ,1},a_{k,f}^{\min ,2} \right\} 
		~\mathrm{and}~a_{k,n}=1-\max \left\{ a_{k,f}^{\min ,1},a_{k,f}^{\min ,2} \right\} , k\in \mathcal{K}.
	\end{equation}
	
For any given $\left\{ \mathbf{a}_k \right\} $ and $\mathbf{V}$, the optimization problem for the design of $\left\{ \mathbf{W}_k \right\} $ is expressed as follows
\begin{subequations}\label{AB_Max_SDP}
  \setlength{\abovedisplayskip}{5pt}
  \setlength{\belowdisplayskip}{5pt}
	\begin{align}
		&\underset{\chi >0,\mathbf{W}_k\succcurlyeq 0}{\max}\chi , \\
		&\mathrm{s}.\mathrm{t}.~\eqref{R11 constrait},~\eqref{R21 constrait},~\eqref{R22 constrait},~\eqref{OP_MM_SDP:b},~\eqref{OP_MM_SDP:d},~\eqref{OP_MM_SDP_smooth:b}. 	 
	\end{align} 
\end{subequations}  	
 
Similar to tackling the rank-one constraint in problem~\eqref{ABP_Max_SDP}, we \textcolor[rgb]{0.00,0.00,0.00}{solve} problem~\eqref{AB_Max_SDP} by employing SDR and removing \textcolor[rgb]{0.00,0.00,0.00}{the} rank-one \textcolor[rgb]{0.00,0.00,0.00}{in} constraint~\eqref{OP_MM_SDP:d}. The tightness of the SDR method for problem~\eqref{AB_Max_SDP} can be proved by the following theorem.
	\begin{theorem}\label{Rank1_only_W}
		The optimal solution $\mathbf{W}_k$ satisfying $\text{rank}\left( \mathbf{W}_k \right) \leqslant 1$ can always be obtained by solving problem~\eqref{AB_Max_SDP} without rank-one constraint.
		
		\textit{Proof}: The proof follows similar arguments as that of \textbf{Theorem~\ref{Rank1 in joint problem} }and is thus omitted for brevity.
	\end{theorem}

The overall proposed IAO algorithm is summarized in~\textbf{Algorithm~\ref{proposed low algorithm}}. Similar to~\textbf{Algorithm~\ref{proposed BCD algorithm}}, the objective value obtained in~\textbf{Algorithm~\ref{proposed low algorithm}} is non-decreasing in each iteration and the proposed algorithm is guaranteed to converge. ~\textbf{Algorithm~\ref{proposed low algorithm}} is computationally efficient as $\left\{ \mathbf{W}_k \right\}  $ in step $4$ are updated by solving the SDP problem~\eqref{AB_Max_SDP} and $\left\{ \mathbf{a}_{k} \right\} $ in step $6$ are updated by using closed-form expressions without \textcolor[rgb]{0.00,0.00,0.00}{iterations}. The computational complexity of problem~\eqref{AB_Max_SDP} in solving the SDP problem can be represented by $\mathcal{O} \left( \max \left\{ 3K+Q+1,N \right\} ^4\sqrt{N}\log \frac{1}{\varpi _5} \right) 
$, where $\varpi _5$ is the corresponding solution accuracy. Thus, the overall complexity of~\textbf{Algorithm~\ref{proposed low algorithm}} can be written as $\mathcal{O} \left( T_{5}^{\max}\left( \max \left\{ 3K+Q+1,N \right\} ^4\sqrt{N}\log \frac{1}{\varpi _5}+O_3 \right) \right) $, where $T_{5}^{\max}$ is the \textcolor[rgb]{0.00,0.00,0.00}{number of iterations} of~\textbf{Algorithm~\ref{proposed low algorithm}}.
	\begin{algorithm}
		\caption{Proposed IAO algorithm for solving problem~\eqref{OP_MM}}
		\label{proposed low algorithm}
		\begin{algorithmic}[1]
			\STATE  Initialize $\left\{ \mathbf{a}_{k}^{\left( 0 \right)} \right\} $ and $\mathbf{V}^{\left( 0 \right)}$ and set the iteration index ${t_5} = 0$.
			\REPEAT
			\STATE  ${t_5} = {t_5} + 1$;
			\STATE  For a given $\mathbf{a}_{k}^{\left( t_5-1 \right)}$ and $\mathbf{V}^{\left( t_5-1 \right)}$, update $\mathbf{W}_{k}^{\left( t_5 \right)}$ by solving problem~\eqref{AB_Max_SDP} without constraint~\eqref{OP_MM_SDP:d} ; 
			\STATE  For a given $\mathbf{W}_{k}^{\left( t_5 \right)}$ and $\mathbf{a}_{k}^{\left( t_5 -1 \right)}$, update $\mathbf{V}^{\left( {t_5 } \right)}$ by applying \textbf{Algorithm~\ref{PB_algorithm}};
			\STATE   For a given $\mathbf{W}_{k}^{\left( t_5 \right)}$ and $\mathbf{V}^{\left( t_2 \right)}$, update $\mathbf{a}_{k}^{\left( t_5 \right)} $ according to~\eqref{power_allocation_sub};
			\UNTIL {the objective value of problem~\eqref{OP_MM_SDP_smooth} converges.}
			\STATE   \textbf{Output}: $\mathbf{W}_{k} $, $\mathbf{a}_{k}$, and $\mathbf{V}$, $ k\in \mathcal{K}$.
		\end{algorithmic}
	\end{algorithm} 
	
 \section{Simulation Results}  
\begin{figure}[t]
 	\setlength{\belowcaptionskip}{-0.6cm}   
 	\centering
 	\includegraphics[scale=0.22]{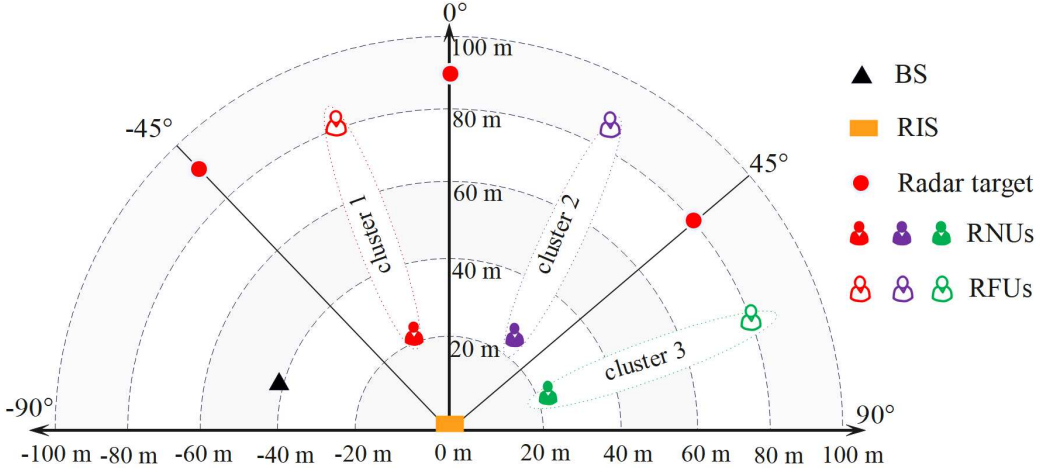}
 	\caption{The simulated RIS-NOMA-ISAC system geometry.}
 	\label{simulation_fig}
 \end{figure}

 In this section, the performance of the proposed algorithms
 for \textcolor[rgb]{0.00,0.00,0.00}{the} RIS-NOMA-ISAC system is evaluated through numerical simulations.
 The simulated RIS-NOMA-ISAC system geometry is shown in Fig.~\ref{simulation_fig}. We assume that there are three clusters and three radar targets. The RIS is located \textcolor[rgb]{0.00,0.00,0.00}{at the origin}, while the BS is located at $\left( -40, 10\right) $ meter $\left( \text{m} \right) $. In each cluster, the RNU and the RFU are randomly distributed on the half circles centered at $\left( 0,0 \right) \text{m}$ with \textcolor[rgb]{0.00,0.00,0.00}{radii} of $r_{k,n}\in \left[ 20,25 \right]\mathrm{m} $ and $r_{k,f}\in \left[ 80,85 \right] \mathrm{m}$, respectively.  Let $\theta _{k,i}$ denote the angle from the RIS to user $\mathrm{U}\left( k,i \right)$ and we further assume that $\left\{ \theta _{1,i} \right\} $, $\left\{ \theta _{2,i} \right\} $, and $\left\{ \theta _{3,i} \right\} $, which denote the set of angles of Cluster $1$, $2$ and $3$, are randomly distributed in the angle ranges $\left( -30^{\text{o}},-20^{\text{o}} \right] $, $\left(20^{\text{o}},30^{\text{o}} \right] 
 $, and $\left( 60^{\text{o}},70^{\text{o}} \right] $, respectively. Without loss of generality, let $\theta _{k,n}=\theta _{k,f}$, $k\in \mathcal{K}$. The angles from the RIS to the three targets are set to be $-45^{\text{o}}$,~$0^{\text{o}}$ and~$45^{\text{o}}$, and the corresponding \textcolor[rgb]{0.00,0.00,0.00}{radii} are set to be $90~\text{m}$, $90~\text{m}$ and $80~\text{m}$, respectively. Given the angles of sensing targets, the desired beampattern can be defined as   
 \begin{equation}\label{desired_beampattern}
 	  \setlength{\abovedisplayskip}{5pt}
 	\setlength{\belowdisplayskip}{5pt}
 	\mathcal{P}\left( \theta \right) =\left\{ \begin{array}{c}
 		1, \theta _{\text{T}}-\frac{\Delta \theta}{2}\leqslant \theta \leqslant \theta _{\text{T}}+\frac{\Delta \theta}{2},\theta _{\text{T}}\in \left\{ -45^{\text{o}},0^{\text{o}},45^{\text{o}} \right\},\\
 		0, \text{otherwise},~~~~~~~~~~~~~~~~~~~~~~~~~~~~~~~~~~~~~~~~~\\
 	\end{array} \right. 
 \end{equation}
 where $\Delta \theta =6^{\text{o}}$ is the desired beam width, $\theta$ is the element of the angle grid $\left[ -90^{\mathrm{o}}:\frac{180^{\mathrm{o}}}{100}:90^{\mathrm{o}} \right]$, and the interested sensing angle set can be defined as $\mathcal{Q} _{\theta}=\left\{ \theta \left| \begin{array}{c}
 	\theta _{\mathrm{T}}-\frac{\Delta \theta}{2}\leqslant \theta \leqslant \theta _{\mathrm{T}}+\frac{\Delta \theta}{2},\\
 	\theta _{\mathrm{T}}\in \left\{ -45^{\mathrm{o}},0^{\mathrm{o}},45^{\mathrm{o}} \right\}\\
 \end{array} \right. \right\} $.
  
  We further assume that all the channels follow the Rician fading \textcolor[rgb]{0.00,0.00,0.00}{distribution}, which can be modeled as follows~\cite{Li2021JointBD,9449661}: 
 \begin{equation}\label{channel_G}
 	  \setlength{\abovedisplayskip}{5pt}
 	\setlength{\belowdisplayskip}{5pt}
 	\mathbf{G}=\sqrt{\varepsilon _0d_{\text{BR}}^{-\varpi _{\text{BR}}}}\left( \sqrt{\frac{\kappa _{\text{BR}}}{1+\kappa _{\text{BR}}}}\mathbf{G}_{\text{LoS}}+\sqrt{\frac{1}{1+\kappa _{\text{BR}}}}\mathbf{G}_{\text{NLoS}} \right), 
 \end{equation}
 \begin{equation}\label{channel_gki}
 	  \setlength{\abovedisplayskip}{5pt}
 	\setlength{\belowdisplayskip}{5pt}
 	\mathbf{g}_{k,i}=\sqrt{\varepsilon _0d_{\text{R},\left( k,i \right)}^{-\varpi _{\text{R},\left( k,i \right)}}}\left( \sqrt{\frac{\kappa _{\text{R},\left( k,i \right)}}{1+\kappa _{\text{R},\left( k,i \right)}}}\mathbf{g}_{\text{R},\left( k,i \right)}^{\text{LoS}}+\sqrt{\frac{1}{1+\kappa _{\text{R},\left( k,i \right)}}}\mathbf{g}_{\text{R},\left( k,i \right)}^{\text{NLoS}} \right), 
 \end{equation}
 where $\varepsilon _0$ denotes the path loss at a reference distance of one meter, $\varpi _{\text{BR}}$ and $\varpi _{\text{R},\left( k,i \right)}$ denote the path loss exponents, $\kappa _{\text{BR}}$ and $\kappa _{\text{R},\left( k,i \right)}$ denote the Rician factors, $\mathbf{G}_{\text{LoS}}$ and $\mathbf{g}_{\text{R},\left( k,i \right)}^{\text{LoS}}$ are the LoS components of channel $\mathbf{G}$ and $\mathbf{g}_{k,i}$, $\mathbf{G}_{\text{NLoS}}$ and $\mathbf{g}_{\text{R},\left( k,i \right)}^{\text{NLoS}}$ are the corresponding NLoS components. The LoS component is modeled as the product of the array response vectors of the transceivers and the NLoS component is modeled as Rayleigh fading. Without loss of generality, we assume that the path loss exponents and the Rician factors for all channels are identical. The path loss at a reference distance of one meter is set to $30{\text {~dB}}$, the path loss exponents are set to 2.2, the Rician factors are set to 3, the noise power is set to $-90\text{~dBm}$, the normalized spacing between two adjacent antennas(elements) is set as $\frac{d}{\lambda}=0.5$. Other system parameters are
 set as follows unless specified otherwise later: the maximum transmit power is set to $35$ dBm, the minimum QoS requirement for RNUs and RFUs are set to $R_{k,n}^{\min}=0.5$ bits/s/Hz and $R_{k,f}^{\min}=0.1$ bits/s/Hz, respectively. Unless otherwise stated, we adopt the default values of the system parameters provided in Table~\ref{parameter}. All \textcolor[rgb]{0.00,0.00,0.00}{of} the following numerical results are obtained
 by averaging over $100$ random channel realizations unless otherwise specified.
 \begin{table}[t]
 	\setlength{\abovecaptionskip}{3pt}
 	\setlength{\belowcaptionskip}{3pt}
 	\caption{Simulation Parameters}
 	\label{parameter}
 	\begin{center}
 		\begin{tabular} {|l|l|}
 			\hline
 			Parameter &  Value \\
 			\hline
 			Locations of the BS and RIS & $\left( -40,10\right) {\rm m} $ and  $\left( 0,0\right) {\rm m}$		\\	
 			\hline
 			Angles of the three radar targets &   $-45^{\rm o}$, $0^{\rm o}$, and $45^{\rm o}$       \\
 			\hline
 			Radius from the RIS to the three radar targets   & $90{\rm m}$, $90{\rm m}$, and $80{\rm m}$	 \\
 			\hline 	
 			Angle ranges of the three clusters &  $\left( -30^{\text{o}},-20^{\text{o}} \right] $, $\left(20^{\text{o}},30^{\text{o}} \right] 
 			$, and $\left( 60^{\text{o}},70^{\text{o}} \right] $ \\
 			\hline 	
 			Radius ranges from the RIS to RNUs and RFUs & 	$ \left[ 20,25 \right]\mathrm{m} $ and $\left[ 80,85 \right] \mathrm{m}$ \\
 			\hline 	
 			Minimum QoS requirement for RNUs and RFUs & $0.5$  bits/s/Hz and $0.1$ bits/s/Hz \\
 			\hline 	
			Maximum transmit power                    & $35$ dBm \\ 			
 			\hline 			 				
 			Path loss exponents and Rician factors of all channels   &  $ 2.2$ and $ 3$   \\ 	
 			\hline
 			Path loss at one meter      &		$30$~{\rm dB}	 \\			
 			\hline
 			Desired beam width   &    $6^{\text{o}}$ \\
 			\hline
 			Noise power     &    \textcolor[rgb]{0.00,0.00,0.00}{$-90$~{\rm dBm}} \\ 			
 			\hline 			
 		\end{tabular}
 	\end{center}
 \end{table}
\subsection{Convergence and Solutions' Rank of the Proposed Algorithms} 
\subsubsection{Convergence}
\begin{figure}[H]
 	\setlength{\belowcaptionskip}{-0.6cm}   
 	\centering 
 	\begin{minipage}[t]{0.48\textwidth}  
 		\centering
 		\includegraphics[scale=0.54]{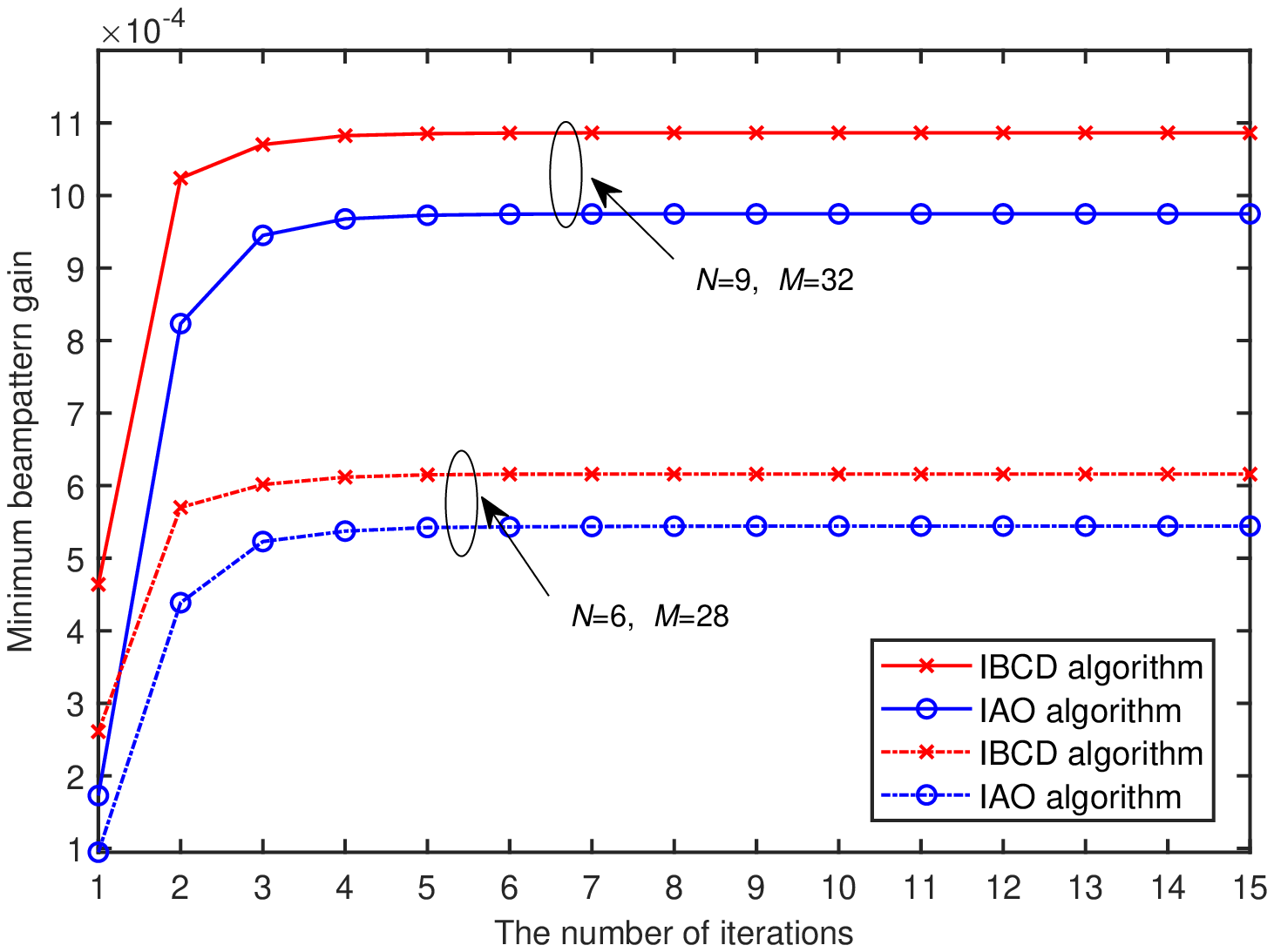}
 		\caption{Convergence of the proposed IBCD algorithm and IAO algorithm.}
 		\label{Converge}
 	\end{minipage}
 	\begin{minipage}[t]{0.48\textwidth}  
	\centering
	\includegraphics[scale=0.54]{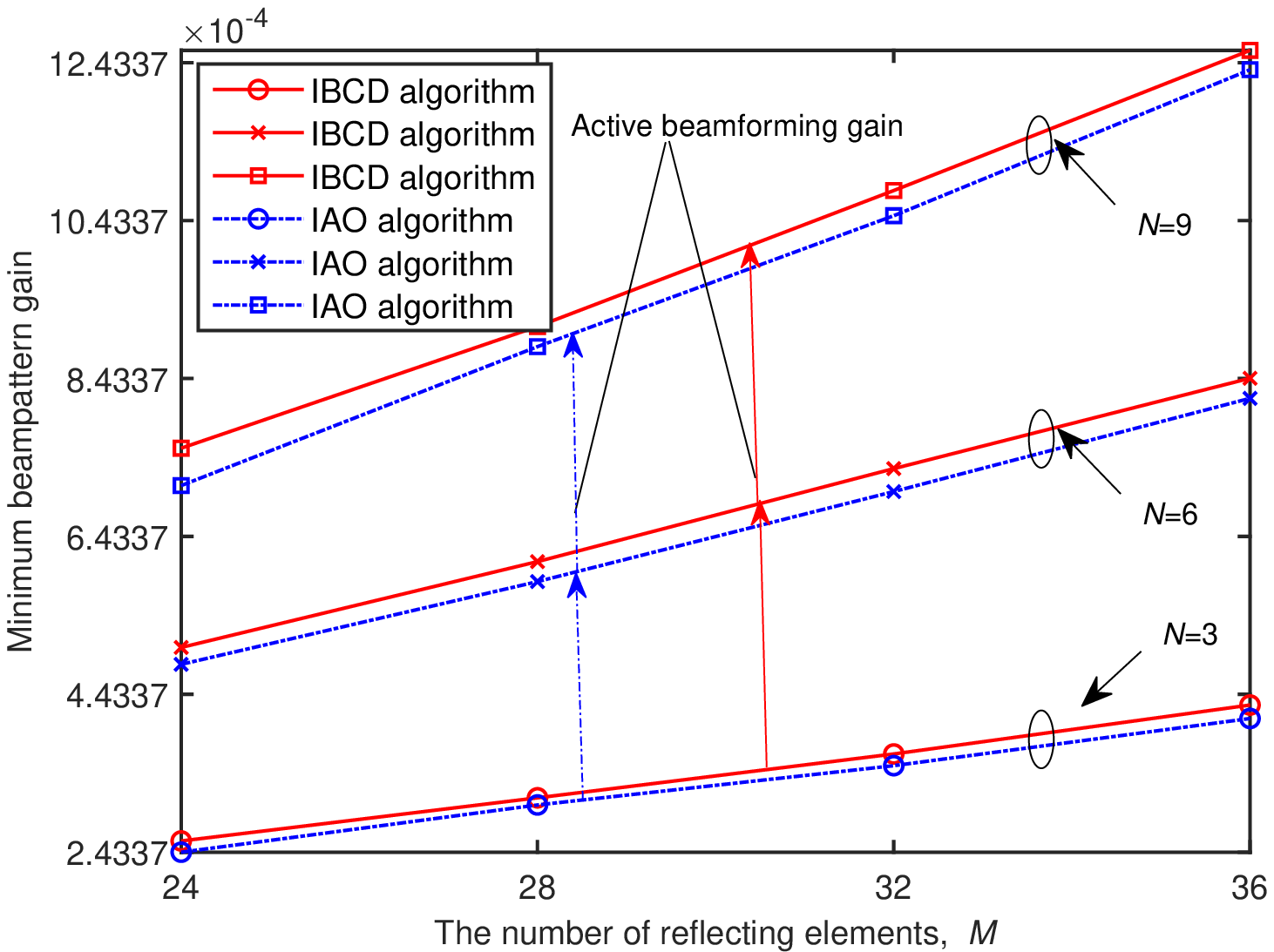}
	\caption{Minimum beampattern gain versus number of reflecting elements for different $N$.}
	\label{MinBPgain_vs_M}
	 	\end{minipage}
 \end{figure}

 In Fig.~\ref{Converge}, we investigate the convergence behavior of the proposed algorithms with different numbers of RIS reflecting elements $M$ and BS antennas $N$.
We observe that, for both cases, the proposed algorithms monotonically converge to stationary points within approximately \textcolor[rgb]{0.00,0.00,0.00}{$10$} iterations.  
Furthermore, the minimum beampattern gain achieved by the IAO algorithm is lower than that of the IBCD algorithm.
 This \textcolor[rgb]{0.00,0.00,0.00}{behavior} will be explained in the following numerical results. Though some performance loss is incurred by the IAO algorithm, the complexity of the IAO algorithm is much lower than that of the IBCD algorithm.
\subsubsection{Solutions' Rank} 
Based on \textbf{Theorem~\ref{Rank1 in joint problem}} and \textbf{Theorem~\ref{Rank1_only_W}}, the rank-one solutions for active beamforming matrices can always be obtained by the proposed IBCD and IAO algorithms. To verify the Theorems, in Table~\ref{ratio_eig}, we list the average ratios between the largest eigenvalue and the second largest eigenvalue of the matrices $\left\{ \mathbf{W}_1,\mathbf{W}_2,\mathbf{W}_3 \right\} $ obtained by the proposed algorithms. It is easy to observe that the \textcolor[rgb]{0.00,0.00,0.00}{values} of the ratios are always enormous \textcolor[rgb]{0.00,0.00,0.00}{for} different \textcolor[rgb]{0.00,0.00,0.00}{values} of $N$ and $M$. 
On the other hand, we also list \textcolor[rgb]{0.00,0.00,0.00}{this} ratio \textcolor[rgb]{0.00,0.00,0.00}{for} the passive beamforming matrix $\mathbf{V}$ obtained by \textbf{Algorithm~\ref{PB_algorithm}} in Table~\ref{ratio_eig}. Obviously, the average ratios achieved by the proposed IBCD and IAO algorithms are sufficiently large. These results in Table~\ref{ratio_eig} reveal that the solutions obtained by the proposed algorithms always satisfy the rank-one constraints.
\begin{table*}[!ht]
	\caption{The ratios obtained by the proposed algorithms with different $N$ and $M$.}
	\label{ratio_eig}
	\center
	\begin{tabular} {|c|c|c|c|c|c|c|c|}  \cline{1-6}		
	 \multirow{6}{*}{ IBCD }     	&     $M$     	&\multicolumn{2}{c|}{$32$}  &   \multicolumn{2}{c|}{$36$}    \\  \cline{2-2} \cline{3-4} \cline{5-6}
		
	 \multirow{6}{*}{algorithm}  	&     $N$     	&   3      &    12    &     3      &     12 	   	\\	 \cline{2-2}  \cline{3-6}
		
	 \multirow{6}{*}{}  	 	    &$\mathbf{W}_1$	& 2.4441e+09 & 6.0565e+09 & 2.0657e+09 & 3.7819e+09   	\\  \cline{2-6} 	

	 \multirow{6}{*}{}  	        &$\mathbf{W}_2$	& 2.5464e+09 & 5.4004e+09 & 2.0341e+09 & 4.1430e+09    	\\	 \cline{2-6} 
		
	 \multirow{6}{*}{}  		  	&$\mathbf{W}_3$	& 2.6132e+09 & 5.9513e+09 & 2.1736e+09 & 4.0795e+09  	\\	\cline{2-6} 	
		 
	 \multirow{6}{*}{} 				&$\mathbf{V} $	& 2.4804e+10 & 3.0715e+10 & 2.5339e+10 & 0.5839e+10  	\\	  \cline{1-6}	
	 		
	  \multirow{6}{*}{IAO } 		&     $M$     	&\multicolumn{2}{c|}{$32$}  	&   \multicolumn{2}{c|}{$36$}    \\  \cline{2-2} \cline{3-4} \cline{5-6}
	 
	 \multirow{6}{*}{algorithm}  	&     $N$     	&           3    &    12       	&     3      &     12 	   	\\	 \cline{2-2}  \cline{3-6}                          
	 
	 \multirow{6}{*}{}  	 		&$\mathbf{W}_1$	& 0.0728e+11 & 1.7223e+11 & 2.0472e+10 & 7.6990e+10 	\\  \cline{2-6} 	 
	 
	 \multirow{6}{*}{}  			&$\mathbf{W}_2$	& 0.1226e+11 & 1.0293e+11 & 1.0012e+10 & 5.9330e+10   	\\	 \cline{2-6} 
	 
	 \multirow{6}{*}{}  			&$\mathbf{W}_3$	& 0.1436e+11 & 1.1674e+11 & 1.3838e+10 & 8.1600e+10 	\\	\cline{2-6} 	
	 
	 \multirow{6}{*}{} 				&$\mathbf{V} $	& 1.2814e+11 & 1.8597e+11 & 5.9715e+11 & 3.7944e+11   	\\	  \cline{1-6}		
	\end{tabular}
\end{table*}
 
 \subsection{Performance Analysis of the Proposed Algorithms} 
Fig.~\ref{MinBPgain_vs_M} depicts the minimum beampattern gain versus the number of RIS reflecting elements, $M$, for different \textcolor[rgb]{0.00,0.00,0.00}{numbers} of BS \textcolor[rgb]{0.00,0.00,0.00}{antennas}. 
Firstly, it is observed that the minimum beampattern gains obtained by the proposed algorithms monotonically increase with $M$. This is expected since installing more reflecting elements at RIS can introduce more virtual LoS links and provide higher passive beamforming gain towards the radar targets. 
Secondly, the proposed low complexity IAO algorithm can achieve comparable performance to that achieved by the IBCD algorithm. 
Thirdly, we note that the performance of the two proposed algorithms can be improved by \textcolor[rgb]{0.00,0.00,0.00}{increasing $N$}. In fact, more BS antennas introduce more DoFs to construct a more directional sensing beam and to achieve a higher active beamforming gain, thereby increasing minimum beampattern gain.

\begin{figure}[H]
    \setlength{\abovecaptionskip}{3pt}
    \setlength{\belowcaptionskip}{-25pt} 
 	\centering 
 	\begin{minipage}[t]{0.48\textwidth}  
	 	 \centering
		\includegraphics[scale=0.54]{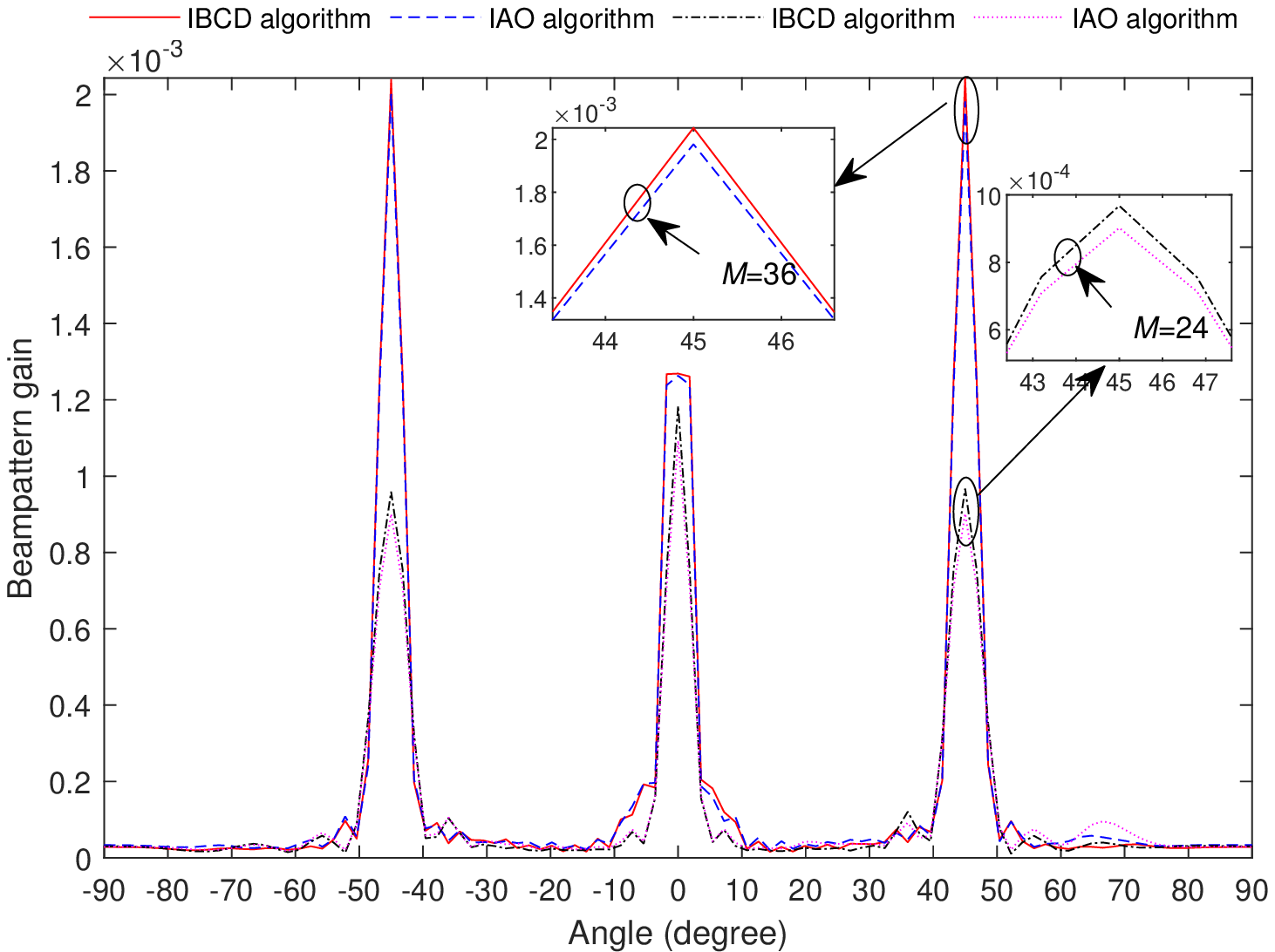}
		\caption{Beampattern gain versus angles for different $M$ with $N=9$.}
		\label{RIS_NOMA_ISAC_BPgain_vs_angle}
	 	\end{minipage}
 	\begin{minipage}[t]{0.48\textwidth}  
 			\centering
	 		\includegraphics[scale=0.54]{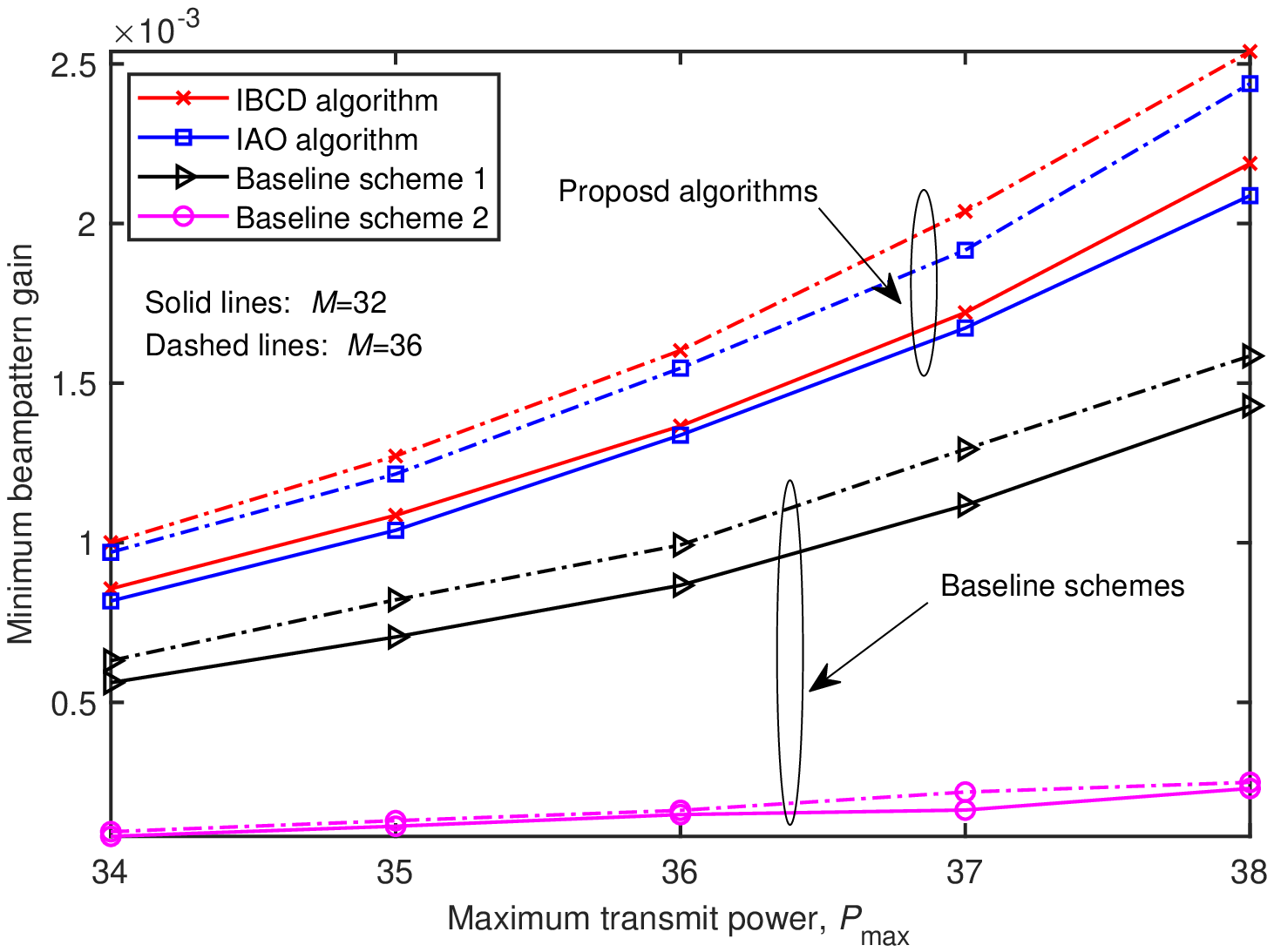}
	 		\caption{Minimum beampattern gain versus maximum transmit power, $P_{\mathrm{max}}$, for different schemes with $N=9$.}
	 		\label{MinBPgain_vs_Pmax}
 	\end{minipage}
 \end{figure}

On the other hand, in Fig.~\ref{RIS_NOMA_ISAC_BPgain_vs_angle}, we plot the beampattern gain versus angles. As illustrated in Fig.~\ref{RIS_NOMA_ISAC_BPgain_vs_angle}, both proposed schemes can achieve the dominant peaks of the beampattern gain in the angles of interest, i.e., $-45^{\text{o}}$,~$0^{\text{o}}$ and~$45^{\text{o}}$.
 Moreover, at the target directions, the achievable beampattern gains of the IBCD algorithm are always higher than that of the IAO algorithm. 
\subsection{Comparison With Different Schemes for RIS-NOMA-ISAC System} 	 
We compare the \textcolor[rgb]{0.00,0.00,0.00}{performances} of the proposed algorithms with two baseline schemes. 
For baseline scheme 1, the active beamforming design implements an approximate zero-forcing method~\cite{Wang2017SpectrumAE}, i.e, only the RNUs in each cluster and the combined channel vector $\mathbf{g}_{k,n}^{H}\mathbf{\Theta G}$ are utilized to determine each beam.
For baseline scheme 2, the active beamforming \textcolor[rgb]{0.00,0.00,0.00}{is realized by} the maximum-ratio transmission method~\cite{Wu2019IntelligentRS} based on the RFUs's combined channel gain. 
We further assume that each cluster shares the same power allocation coefficients in all baseline schemes. \textcolor[rgb]{0.00,0.00,0.00}{For a fair comparison}, \textcolor[rgb]{0.00,0.00,0.00}{the resulting} optimization problems are solved by applying the SDR or SRCR algorithm~\cite{Cao2017ASC}.
In Fig.~\ref{MinBPgain_vs_Pmax}, we study the minimum beampattern gain versus the maximum transmit power, $P_{\max}$, for all schemes. 
As expected, the minimum beampattern gain increases as $P_{\max}$ grows. The reason behind this is that when the transmit power is high, the received signal strength at the communication users and targets are strong. The QoS requirements of the communication users \textcolor[rgb]{0.00,0.00,0.00}{can be easily} satisfied and more redundant power can be utilized to improve the sensing performance, thereby significantly increasing the beampattern gain. Moreover, it can be seen that the proposed schemes are capable of \textcolor[rgb]{0.00,0.00,0.00}{providing a higher} beampattern gain than the baseline schemes. This can be explained by the fact that the active beamforming, power allocation coefficients and passive beamforming are not jointly optimized in the baseline schemes. However, in the proposed schemes, the joint optimization over these variables can provide a considerable performance improvement and fully exploit the DoFs introduced by the RIS and NOMA.
  
\subsection{Comparison With Different RIS Assisted Systems}  
\textcolor[rgb]{0.00,0.00,0.00}{To} demonstrate the effectiveness of \textcolor[rgb]{0.00,0.00,0.00}{our} proposed RIS-NOMA-ISAC system, we consider the following two benchmark systems: 
\begin{itemize}
	\item  RIS-ISAC system: In this system, NOMA is not employed. The achievable rate of user $k$ can be obtained by: $R_k=\log _2\left( 1+\frac{\left| \mathbf{g}_{k}^{H}\mathbf{\Theta Gw}_k \right|^2}{\sum_{i\ne k}^{2K}{\left| \mathbf{g}_{i}^{H}\mathbf{\Theta Gw}_k \right|^2}+\sigma ^2} \right) 
	$, where $\mathbf{g}_k$ is the channel coefficients from the RIS to user $k$, $k\in \left\{ 1,2,\cdots ,2K \right\} $.
	\item RIS-Sensing system: In this system, the transmitted signal at the BS is used only for radar sensing while the communication function is not considered during the system design. 	
\end{itemize} 

\textcolor[rgb]{0.00,0.00,0.00}{Note that the resulting optimization problems} for RIS-ISAC system and RIS-Sensing system can be solved in a similar way as solving problem~\eqref{OP_MM}. In addition, for the ease of presentation, we focus on the proposed IBCD algorithm for RIS-NOMA-ISAC system.

\begin{figure}[H]
    \setlength{\abovecaptionskip}{3pt}
   \setlength{\belowcaptionskip}{-25pt} 
	\centering 
	\begin{minipage}[t]{0.48\textwidth}  
		\centering
		\includegraphics[scale=0.54]{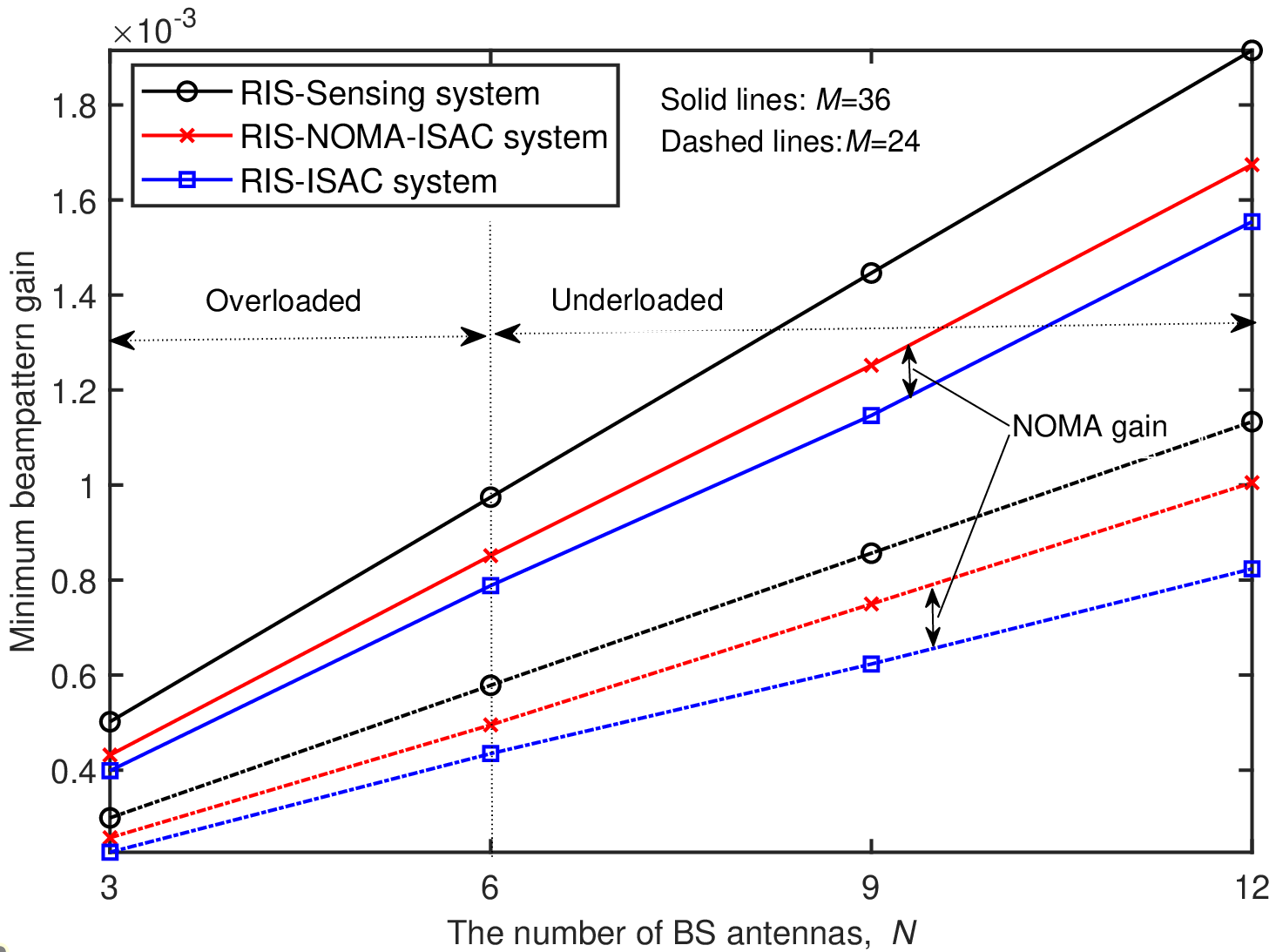}
		\caption{Minimum beampattern gain versus number of BS antennas for different systems.}
		\label{MinBPgain_vs_N}
	\end{minipage}
	\begin{minipage}[t]{0.48\textwidth}  
		\centering
		\includegraphics[scale=0.54]{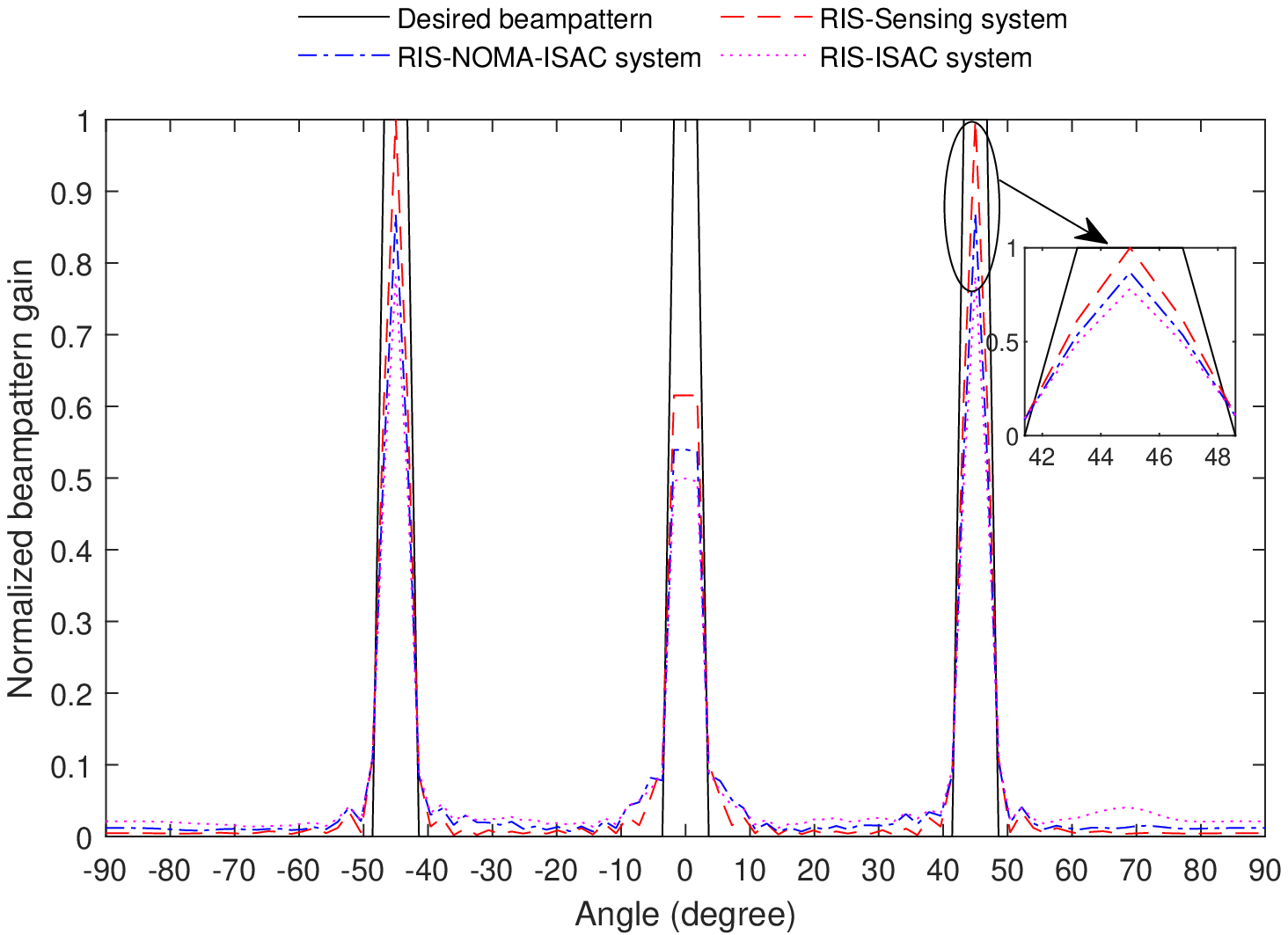}
		\caption{Normalized beampattern gain versus angles for different systems with ${N}=9$ and ${M}=36$.}
		\label{Diff_system_BPgain_vs_angle}
	\end{minipage}
\end{figure}

In Fig.~\ref{MinBPgain_vs_N}, we compare the minimum beampattern gain versus the number of BS antennas, $N$, for different systems. It is clear that \textcolor[rgb]{0.00,0.00,0.00}{when $N$ increases}, which means that more active beams are exploited to transmit the BS signal, higher possible beamforming gain can be provided towards the radar targets.
Moreover, the proposed RIS-NOMA-ISAC system outperforms the RIS-ISAC system under overloaded ($N<6$) and underloaded ($N>6$) cases. This is because the RIS-NOMA-ISAC system can mitigate the inter-user interference by employing SIC and provide more DoFs for radar sensing. However, \textcolor[rgb]{0.00,0.00,0.00}{the RIS-ISAC system cannot mitigate the inter-user interference effectively}. 
Besides, the RIS-NOMA-ISAC system and the RIS-ISAC system achieve lower performance than the RIS-Sensing system, which reveals a trade-off between the radar sensing and  communication for ISAC systems. 

On the other hand, we evaluate the normalized beampattern gain of \textcolor[rgb]{0.00,0.00,0.00}{the} considered systems with respect to different angles in Fig.~\ref{Diff_system_BPgain_vs_angle}. As can be observed, the beampattern gains of all systems have peaks towards target directions and our proposed RIS-NOMA-ISAC system has a stronger peak compared with the RIS-ISAC system. Moreover, the proposed RIS-NOMA-ISAC system can obtain higher beampattern gains at the worst angles than \textcolor[rgb]{0.00,0.00,0.00}{the} RIS-ISAC system. 
\subsection{Illumination Power} 

\begin{figure*}[t!]
    \setlength{\abovecaptionskip}{5pt}
 	\setlength{\belowcaptionskip}{-35pt} 
	\centering
 	\subfigure[RIS-Sensing system]
	{
		\begin{minipage}[t]{0.32\linewidth}	  
			\label{Sensing_N12_M36_IP}
			\includegraphics[scale=0.37]{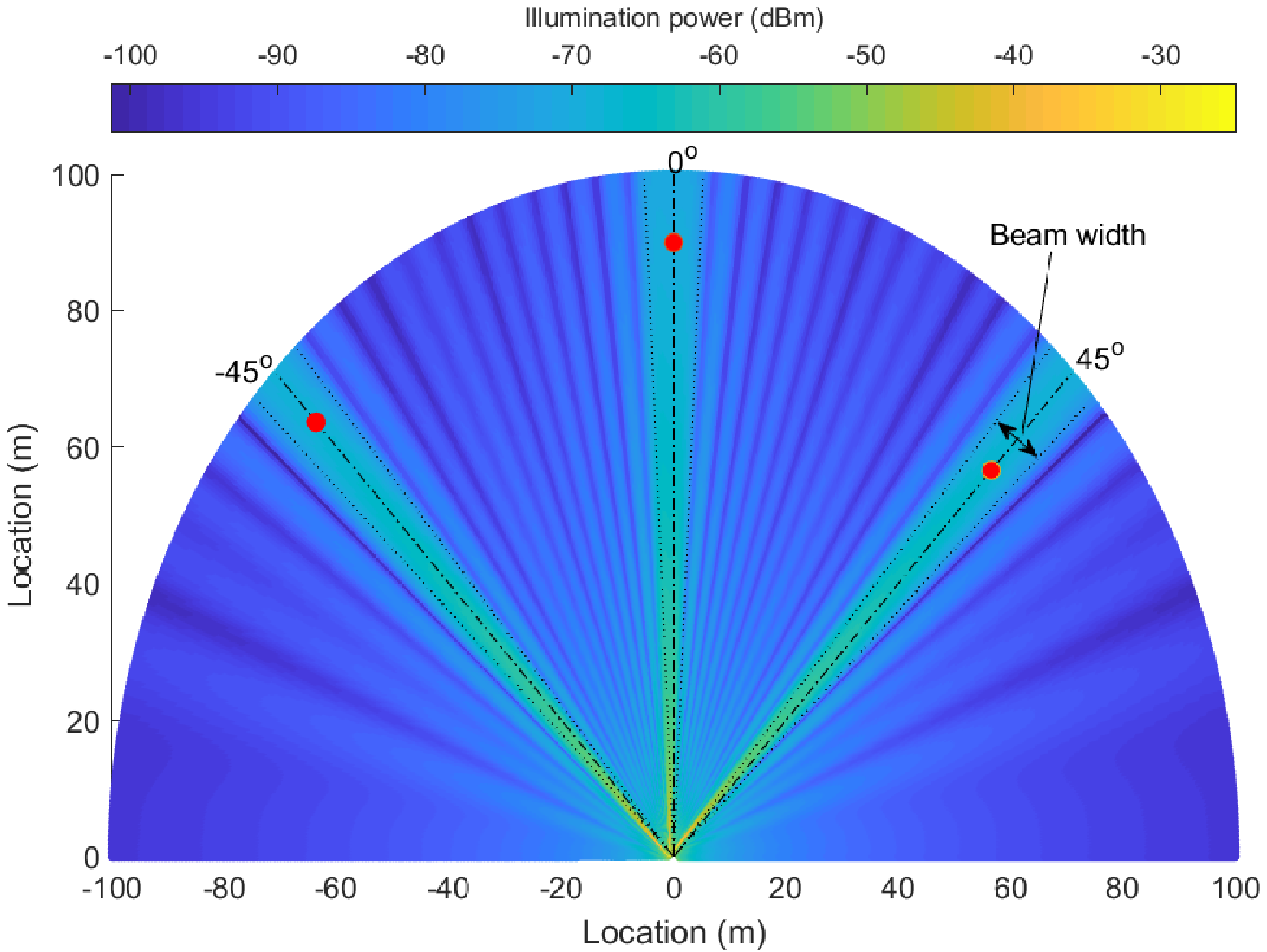}
		\end{minipage}%
	}%
 	\centering
	\subfigure[RIS-NOMA-ISAC system]
	{
		\begin{minipage}[t]{0.32\linewidth}
			\label{RIS_NOMA_ISAC_N12_M36_IP}
			\includegraphics[scale=0.36]{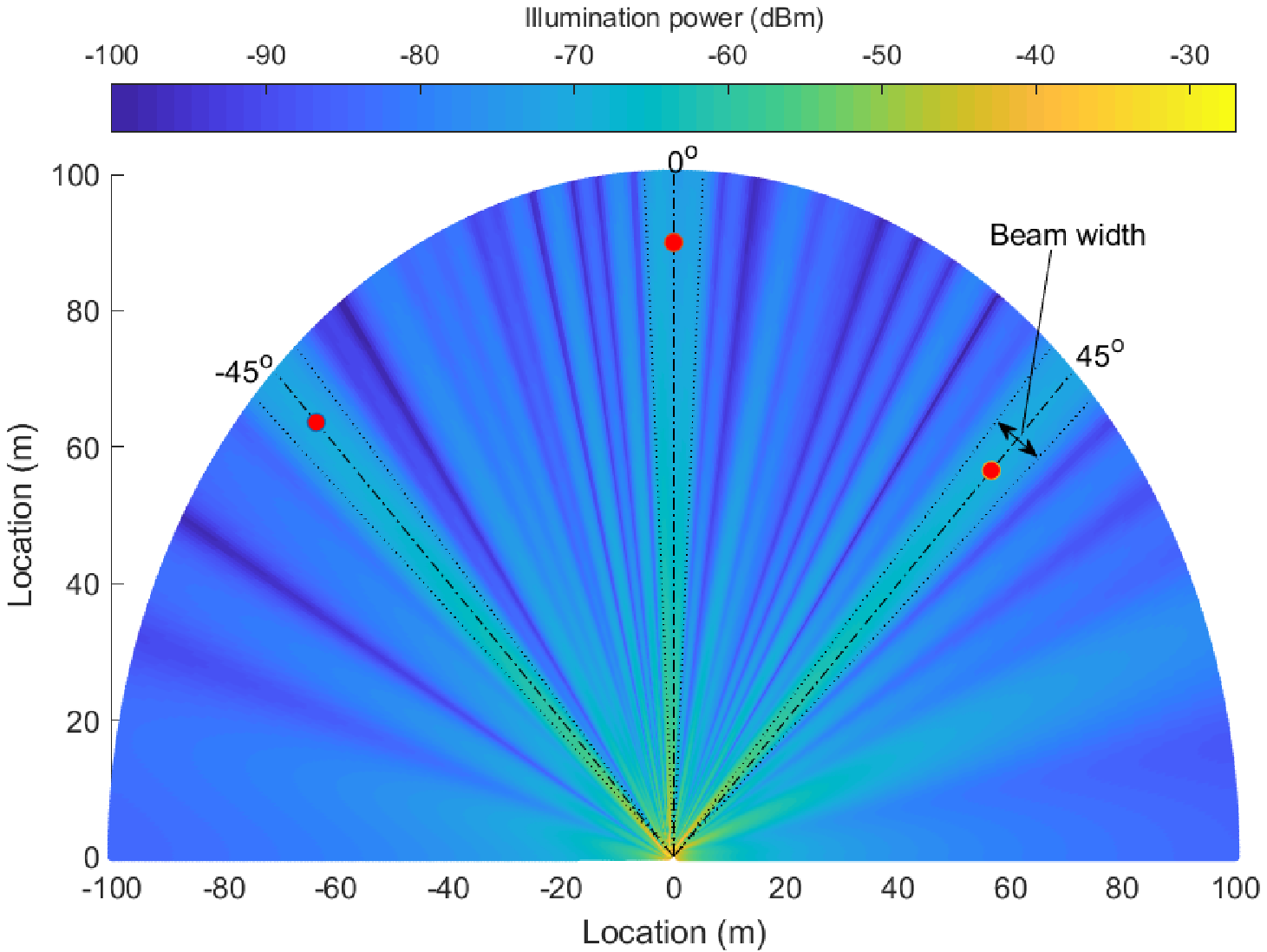}
		\end{minipage}%
	}
	\centering
	\subfigure[RIS-ISAC system]
	{
		\begin{minipage}[t]{0.32\linewidth}
			\label{RIS_ISAC_N12_M36_IP}
			\includegraphics[scale=0.36]{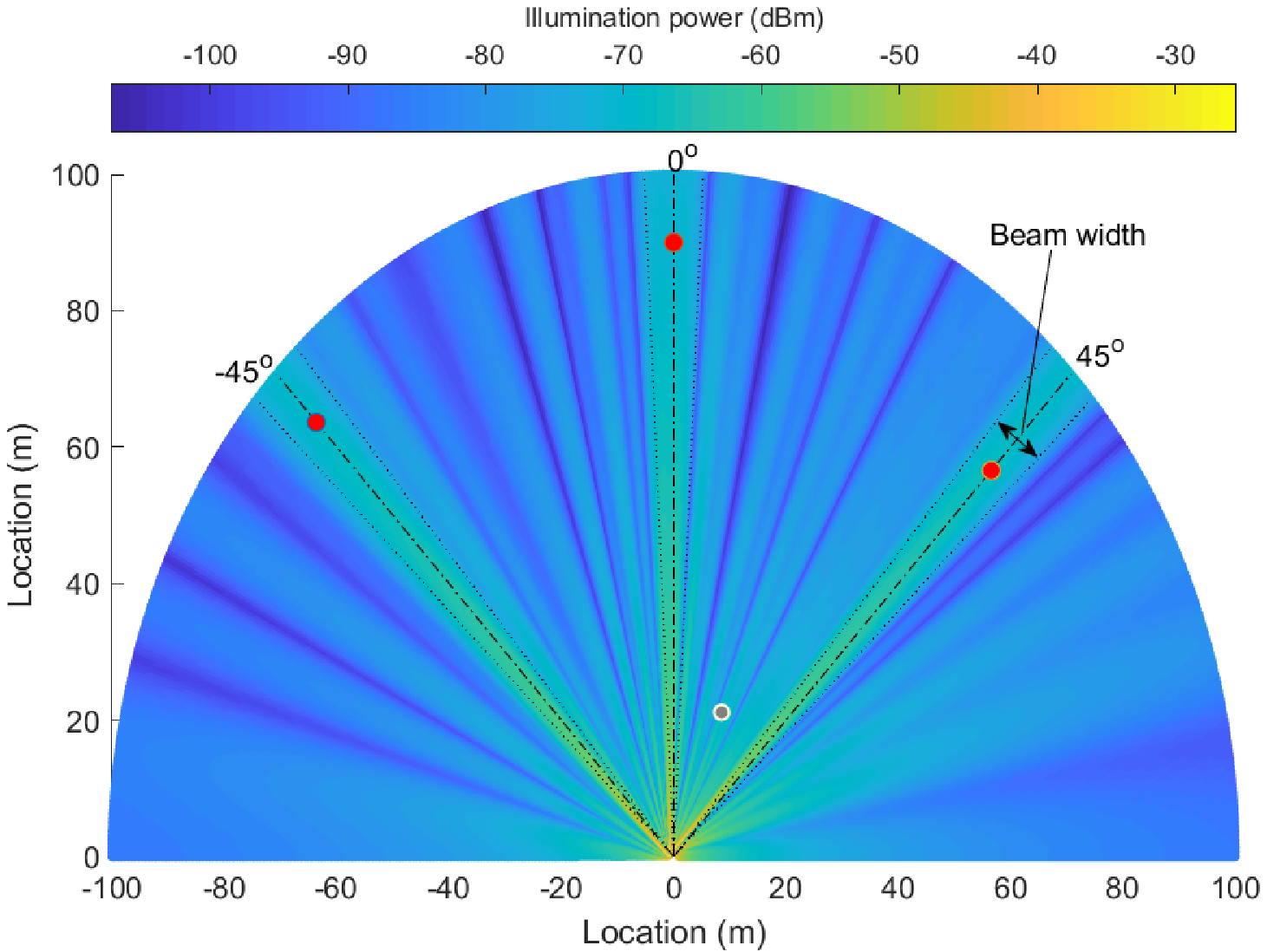}
		\end{minipage}%
	}
  \caption{The map of illumination power for different systems with $N=12$ and $M=36$.}
  \label{illumation_power_map}
\end{figure*}

\begin{figure}[H]
      \setlength{\abovecaptionskip}{5pt}
	 \setlength{\belowcaptionskip}{5pt}
	\centering
	\includegraphics[scale=0.6]{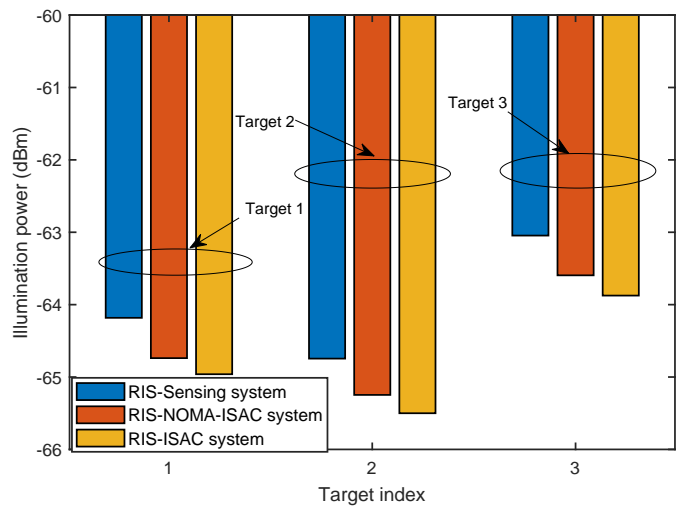}
	\caption{Illumination power on targets for different systems with $N=12$ and $M=36$.}
	\label{illumate_power_target}
\end{figure}

In \textcolor[rgb]{0.00,0.00,0.00}{the} previous subsections, the performance of the proposed schemes and system are evaluated by the beampattern gain metric. In this subsection, we adopt \textcolor[rgb]{0.00,0.00,0.00}{an} alternative metric, namely, illumination power~\cite{Sankar2022BeamformingII}, to evaluate the radar sensing performance. Let $\mathbf{g}_{\mathrm{X}}$ denote the channel from the RIS to any location in the considered system geometry. Then, the illumination power can be defined as $\mathrm{Tr}\left[ \mathbf{V\Upsilon }\left( \sum_{k=1}^K{\mathbf{W}_k} \right) \mathbf{\Upsilon }^H \right]$, where $\mathbf{\Upsilon }=\mathrm{diag}\left\{ \mathbf{g}_{\mathrm{X}}^{H} \right\} \mathbf{G}$.
$\mathbf{g}_{\mathrm{X}}$ is modeled as \textcolor[rgb]{0.00,0.00,0.00}{a} LoS channel as in~\cite{Sankar2022BeamformingII}. In Fig.~\ref{illumation_power_map}, we show the illumination power map of the RIS on different angles and locations over one random channel realization. From Fig.~\ref{illumation_power_map}, we can see that there are three brightest \textcolor[rgb]{0.00,0.00,0.00}{regions} towards the three target directions in different systems. This is intuitive since the RIS attempts to steer both the active and passive beamforming towards the targets to maximize the minimum beampattern. Compared with Fig.\ref{Sensing_N12_M36_IP} for RIS-Sensing system, the leakages of illumination power towards \textcolor[rgb]{0.00,0.00,0.00}{angles of no interest} are more \textcolor[rgb]{0.00,0.00,0.00}{serious} in Fig.~\ref{RIS_NOMA_ISAC_N12_M36_IP} for \textcolor[rgb]{0.00,0.00,0.00}{the} RIS-NOMA-ISAC system and Fig.~\ref{RIS_ISAC_N12_M36_IP} for \textcolor[rgb]{0.00,0.00,0.00}{the} RIS-ISAC system. To further reveal the insights of the illumination power map, in Fig.~\ref{illumate_power_target}, we present the total illumination power towards the \textcolor[rgb]{0.00,0.00,0.00}{angles of interest} for each target, where the \textcolor[rgb]{0.00,0.00,0.00}{angles of interest for} the three targets are $\left[ -45^{\text{o}}-\Delta \theta ,-45^{\text{o}}+\Delta \theta \right] $, $\left[ 0^{\text{o}}-\Delta \theta ,0^{\text{o}}+\Delta \theta \right] 
$ and $\left[ 45^{\text{o}}-\Delta \theta ,45^{\text{o}}+\Delta \theta \right] $, respectively. It is clear that the RIS-Sensing system achieves the best performance. Furthermore, the proposed RIS-NOMA-ISAC system achieves higher illumination power \textcolor[rgb]{0.00,0.00,0.00}{towards} the three targets than that of the RIS-ISAC system. These results clearly demonstrate the importance of employing NOMA in the ISAC system.
 \vspace{-1cm}
 \vspace{-1cm}
 \section{CONCLUSIONS}  
 \vspace{-0.2cm}
 \textcolor[rgb]{0.00,0.00,0.00}{We proposed} a novel RIS-NOMA-ISAC system, where the RIS is deployed to serve the communication users and to assist radar sensing. A joint optimization problem over active beamforming, power allocation coefficients, and passive beamforming was formulated with the aim of maximizing the minimum beampattern gain. Due to the nonconvex nature of the formulated problem, \textcolor[rgb]{0.00,0.00,0.00}{an} IBCD algorithm was proposed to solve the original problem. To further reduce the complexity of the proposed algorithm, a low complexity IAO algorithm was proposed where the exact power allocation coefficients \textcolor[rgb]{0.00,0.00,0.00}{are derived in closed-form expressions}. Numerical results confirmed that \textcolor[rgb]{0.00,0.00,0.00}{our} proposed algorithms can achieve better beampattern gain performance in comparison to other baseline schemes. Moreover, our results revealed that NOMA is an efficient means to mitigate the inter-user interference in ISAC systems \textcolor[rgb]{0.00,0.00,0.00}{whether} the system is overloaded or underloaded. 
  
 
\section*{Appendix~A: Proof of Theorem~\ref{Rank1 in joint problem}} \label{Proof_rank1}
\renewcommand{\theequation}{A.\arabic{equation}}
\setcounter{equation}{0} 

Since problem~\eqref{AB_Max_SDR} is convex and satisfies the Slater's condition, and strong duality holds. The Lagrangian function of problem~\eqref{AB_Max_SDR} in terms of beamforming matrices $\left\{ \mathbf{W}_k \right\} $ is given as follows:
\begin{equation} \label{Lagrangian}  
  \setlength{\abovedisplayskip}{5pt}
\setlength{\belowdisplayskip}{5pt}
	\begin{split}
		\psi _{\text{Lag}} =
		& -\alpha _0\sum_{k=1}^K{\mathrm{Tr}\left( \mathbf{W}_k \right)}-\sum_{k=1}^K{\beta _{k}^{\eta}r_{k,n}^{\min}I_{k,n}^{\mathrm{iter}}} + \sum_{k=1}^K{\mathrm{Tr}\left( \mathbf{W}_k\mathbf{X}_k \right)}+\varrho 	
		\\
		& +\sum_{q=1}^Q{\alpha _q\mathrm{Tr}\left[ \left( \sum_{k=1}^K{\mathbf{W}_k} \right) \mathbf{\Upsilon }_{q}^{H}\mathbf{V\Upsilon }_q \right]}+\sum_{k=1}^K{\mathrm{Tr}\left( \mathbf{Y}_k\left[ \begin{matrix}
				a_{k,n}&		\eta _k\\
				\eta _k&		\mathrm{Tr}\left( \mathbf{W}_k\mathbf{H}_{k,n} \right)\\
			\end{matrix} \right] \right)}
		\\ 
		& +\sum_{k=1}^K{\frac{\beta _{k}^{\mathcal{T} _1}\left( \mathrm{Tr}\left( \mathbf{W}_k\mathbf{\Gamma }_{k,n}^{H}\mathbf{V\Gamma }_{k,n} \right) -r_{k,f}^{\min}I_{k,n}^{\mathrm{iter}} \right)}{\left( 1+r_{k,f}^{\min} \right)}}+\sum_{k=1}^K{\frac{\beta _{k}^{\mathcal{T} _2}\left( \mathrm{Tr}\left( \mathbf{W}_k\mathbf{\Gamma }_{k,f}^{H}\mathbf{V\Gamma }_{k,f} \right) -r_{k,f}^{\min}I_{k,f}^{\mathrm{iter}} \right)}{\left( 1+r_{k,f}^{\min} \right)}},	
	\end{split} 
\end{equation} 
where $\varrho$ is the collection of terms that are not relevant for the proof, $\alpha _0$, $\beta _{k}^{\eta}$, $\alpha _q $, $\beta _{k}^{\mathcal{T} _1}$ and $\beta _{k}^{\mathcal{T} _2}$
are the Lagrange multipliers. The matrix $\mathbf{Y}_k $ and $\mathbf{X}_k $ are the Lagrange multiplier matrix for the positive semi-definite constraints. Note that there exists at least one $\alpha _0 > 0$, since constraint~\eqref{OP_MM_SDP:b} is active for the optimal $\mathbf{W}_k$\footnote{It is easy to prove that constraint~\eqref{OP_MM_SDP:b} is active by \textcolor[rgb]{0.00,0.00,0.00}{the} contradiction method. Specifically, if the equality in \textcolor[rgb]{0.00,0.00,0.00}{constraint}~\eqref{OP_MM_SDP:b} is not satisfied, i.e, $\sum_{k=1}^K{\mathrm{Tr}\left( \mathbf{W}_k \right)}<P_{\max}$. Define $\rho =\frac{P_{\max}}{\sum_{k=1}^K{\mathrm{Tr}\left( \mathbf{W}_k \right)}}>1$ and multiply $\rho$ \textcolor[rgb]{0.00,0.00,0.00}{by} the optimal $\mathbf{W}_k$. As a result, a new solution $\overline{\mathbf{W}}_k$ is obtained. Since the beampattern gain with $\overline{\mathbf{W}}_k$ is larger than that with ${\mathbf{W}}_k$, which contradicts \textcolor[rgb]{0.00,0.00,0.00}{with} the optimality of $\mathbf{W}_k$. Thus, the equality in constraint~\eqref{OP_MM_SDP:b} always holds.}.

The Karush-Kuhn-Tucker (KKT) conditions of problem~\eqref{AB_Max_SDR} that are relevant to the proof are given as follows:
\begin{equation} \label{KKT}
	  \setlength{\abovedisplayskip}{5pt}
	\setlength{\belowdisplayskip}{5pt}
	 \mathrm{K}1:\alpha _0,\alpha _q,\beta _{k}^{\eta},\beta _{k}^{\mathcal{T} _1},\beta _{k}^{\mathcal{T} _2}\geqslant 0,\mathbf{Y}_k,\mathbf{Z}_k\succeq \mathbf{0},~~~\mathrm{K}2:\mathbf{W}_k\mathbf{X}_k=\mathbf{0},~~~\mathrm{K}3:\triangledown _{\mathbf{W}_k}\psi _{\mathrm{Lag}}=\mathbf{0}.
\end{equation}

To proceed, we derive the gradient of $\psi _{\mathrm{Lag}}$ explicitly and rewrite K3 as
\begin{equation} \label{K3}
	  \setlength{\abovedisplayskip}{5pt}
	\setlength{\belowdisplayskip}{5pt}
	\mathbf{Z}_k=\alpha _0\mathbf{I}-\mathbf{\Pi }_k,
\end{equation}
where the matrix $\mathbf{\Pi }_k$ is defined as
\begin{equation} \label{matrix Q}
	  \setlength{\abovedisplayskip}{5pt}
	\setlength{\belowdisplayskip}{5pt}
	\begin{split}
		\mathbf{\Pi }_k 
		&=\left[ \gamma _k\left( r_{k,f}^{\min}a_{k,n}-a_{k,f} \right) +\eta _k\left( r_{k,f}^{\min}a_{k,n}-a_{k,f} \right) -\beta _ka_{k,n} \right] \mathbf{H}_{k,n}^{H}
		\\
		&+\sum_{\widetilde{k}\ne k}^K{\left( \beta _{\widetilde{k}}r_{\widetilde{k},n}^{\min}+\gamma _{\widetilde{k}}r_{\widetilde{k},f}^{\min}+\eta _{\widetilde{k}}r_{\widetilde{k},f}^{\min} \right) \mathbf{H}_{\widetilde{k},n}^{H}}-\sum_{q=1}^Q{\mu _q\mathbf{H}_{q}^{H}}.
	\end{split}
\end{equation}

Denote by $\lambda _{\max}\left( \mathbf{\Pi }_k \right) 
$ the largest eigenvalue of matrix $\mathbf{\Pi }_k$. Due to the randomness of the channels, the largest eigenvalue $\lambda _{\max}\left( \mathbf{\Pi }_k \right) $ is unique. Recalling the expression~\eqref{K3}, if $\lambda _{\max}\left( \mathbf{\Pi }_k \right) >\alpha _0$, then we have: $\alpha _0\mathbf{I}-\mathbf{\Pi }_k\prec \mathbf{0}$ which contradicts \textcolor[rgb]{0.00,0.00,0.00}{with} $\mathbf{Z}_k\succeq \mathbf{0}$. In addition, if $\lambda _{\max}\left( \mathbf{\Pi }_k \right) \leqslant \alpha _0$, we have: $\alpha _0\mathbf{I}-\mathbf{\Pi }_k\succeq \mathbf{0}
$, which implies that $\mathbf{Z}_k \leqslant \mathbf{0}$ and $\mathrm{rank}\left( \mathbf{Z}_k \right) \geqslant N_{\mathrm{T}}-1$. Considering K2, we have: $\mathrm{rank}\left( \mathbf{W}_k \right) \leqslant 1$. The proof is completed.
 
 \section*{Appendix~B: Proof of Theorem~\ref{feaible_check_theorem}} \label{Proof_feaible}
 \renewcommand{\theequation}{B.\arabic{equation}}
 \setcounter{equation}{0} 
 
 According to the QoS constraint~\eqref{R11 constrait} and~\eqref{OP_MM:d}, we have the following inequalities:
 \begin{equation}\label{alpha_kf_max}
  \setlength{\abovedisplayskip}{5pt}
\setlength{\belowdisplayskip}{5pt}
 	a_{k,f}\leqslant 1-\frac{r_{k,n}^{\min}\left[ \sum_{\widetilde{k}\ne k}^K{\text{Tr}\left( \mathbf{W}_{\widetilde{k}}\mathbf{H}_{k,n} \right)}+\sigma ^2 \right]}{\text{Tr}\left( \mathbf{W}_k\mathbf{H}_{k,n} \right)}\triangleq a_{k,f}^{\max}. 
 \end{equation} 
 
 In addition, the QoS constraints~\eqref{R21 constrait} and~\eqref{R22 constrait}  can be respectively rewritten as:
 \begin{equation}\label{alpha_kf_min1}
  \setlength{\abovedisplayskip}{5pt}
\setlength{\belowdisplayskip}{5pt}
 	a_{k,f}\geqslant \frac{\text{Tr}\left( \mathbf{W}_k\mathbf{H}_{k,n} \right) +\sum_{\widetilde{k}\ne k}^K{\left( \mathbf{W}_{\widetilde{k}}\mathbf{H}_{k,n} \right)}+\sigma ^2}{\left( 1+\frac{1}{r_{k,f}^{\min}} \right) \text{Tr}\left( \mathbf{W}_k\mathbf{H}_{k,n} \right)}\triangleq a_{k,f}^{\min ,1},
 \end{equation} 
 \begin{equation}\label{alpha_kf_min2}
  \setlength{\abovedisplayskip}{5pt}
\setlength{\belowdisplayskip}{5pt}
 	a_{k,f}\geqslant \frac{\text{Tr}\left( \mathbf{W}_k\mathbf{H}_{k,f} \right) +\sum_{\widetilde{k}\ne k}^K{\text{Tr}\left( \mathbf{W}_{\widetilde{k}}\mathbf{H}_{k,f} \right)}+\sigma ^2}{\left( 1+\frac{1}{r_{k,f}^{\min}} \right) \text{Tr}\left( \mathbf{W}_k\mathbf{H}_{k,f} \right)}\triangleq a_{k,f}^{\min ,2}.
 \end{equation} 
 
 \textcolor[rgb]{0.00,0.00,0.00}{Combing}~\eqref{alpha_kf_max},~\eqref{alpha_kf_min1} and~\eqref{alpha_kf_min2}, it is easy to observe that if problem~\eqref{OP_MM_SDP_smooth} is feasible, the following condition should be satisfied:
 \begin{equation}\label{max>min}
 	\max \left( a_{k,f}^{\min ,1},a_{k,f}^{\min ,2} \right) \leqslant a_{k,f}^{\max}<1.
 \end{equation}
 

 \vspace{-0.5cm}
\bibliographystyle{IEEEtran}
 \bibliography{zjkbib}

\end{document}